\documentclass[unsortedaddress,reprint]{revtex4-1}
\usepackage{amssymb,latexsym}
\usepackage{amsbsy} % for \boldsymbol
\usepackage{amsmath,amsthm}
\usepackage{mathrsfs}

\usepackage{bm}

\usepackage[shortlabels]{enumitem}
\usepackage{hyperref}

\usepackage{graphicx}
\usepackage{color}

\newcommand{\abs}[1]{\lvert#1\rvert}
\newcommand{\set}[1]{\{ #1 \}}
\theoremstyle{plain}

\theoremstyle{remark}

\theoremstyle{definition}

\newcommand \commentout[1] {}

\begin{document}

%%%%%%%%%%%%%%%%%%%%%%%%%%%%%%%%%%%%%%%%%%%%%%%%
%            Header			                   %
%%%%%%%%%%%%%%%%%%%%%%%%%%%%%%%%%%%%%%%%%%%%%%%%
%UiO Group
\author{Fabian M. Faulstich}
\email{f.m.faulstich@kjemi.uio.no}
\author{Andre Laestadius}
\author{Simen Kvaal}
\affiliation{Hylleraas Centre for Quantum Molecular Sciences, Department of Chemistry, University of Oslo, P.O. Box 1033 Blindern, N-0315 Oslo, Norway}

%Wigner Group
\author{Mih\'aly M\'at\'e, Mih\'aly Andr\'as Csirik}
\author{\"Ors Legeza}
\affiliation{
	%Strongly Correlated Systems "Lend\"ulet" Research Group, Wigner Research Center for Physics, H-1525, Budapest, Hungary
	Wigner Research Center for Physics, H-1525, P.O. Box 49, Budapest, Hungary }

%Prague Group
\author{Libor Veis, Andrej Antalik, Ji\v{r}\'i Brabec}
\author{Ji\v{r}\'i Pittner }
\affiliation{
	J. Heyrovsk\'y Institute of Physical Chemistry, Academy of Sciences of the Czech Republic, v.v.i., Dolej\v{s}kova 3, 18223 Prague 8, Czech Republic
}

%Reinhold
\author{Reinhold Schneider}
\affiliation
{Modeling, Simulation and Optimization in Science, Department of Mathematics, Technische Universit\"at Berlin, Sekretariat MA 5-3, Stra\ss e des 17. Juni 136, 10623 Berlin,Germany}

%Title
\title{Numerical and Theoretical Aspects of the DMRG-TCC Method Exemplified by the Nitrogen Dimer}
%%%%%%%%%%%%%%%%%%%%%%%%%%%%%%%%%%%%%%%%%%%%%%%%
%            Science		                   %
%%%%%%%%%%%%%%%%%%%%%%%%%%%%%%%%%%%%%%%%%%%%%%%%

\begin{abstract}
In this article, we investigate the numerical and theoretical aspects of the coupled-cluster method tailored by matrix-product states. 
We investigate chemical properties of the used method, such as energy size extensivity and the equivalence of linked and unlinked formulation. 
The existing mathematical analysis is here elaborated in a quantum chemical framework. 
In particular, we highlight the use of a so-called CAS-ext gap describing the basis splitting between the complete active space and the external part. 
Moreover, the behavior of the energy error as a function of the optimal basis splitting is discussed.
We show numerical investigations on the robustness with respect to the bond dimensions of the single orbital entropy and the mutual information, which are quantities that are used to choose the complete active space.
Furthermore, we extend the mathematical analysis with a numerical study on the complete active space dependence of the error. 
\end{abstract}

\maketitle

\section*{I. Introduction}
The coupled-cluster (CC) theory has played a revolutionary role in establishing a new level of high accuracy in electronic structure calculations and quantum-chemical simulations. 
Despite the immense progress made in the field, computational schemes aiming at describing quasi-degenerate electronic structures of chemical systems are still unreliable. 
These multi-configuration systems, also called strongly correlated systems, form one of the most challenging computational problems in quantum chemistry. 
Since these systems appear in various research areas, a reliable computational scheme is of major interest for natural sciences. 
Recently, the computational advantages of the novel density matrix renormalization group tailored coupled-cluster (DMRG-TCC) method restricted to single (S) and double (D) excitations were demonstrated on large and statically correlated systems by L.V. {\it et al.} in Refs.~\cite{veis2016coupled,veis2018intricate}.
Furthermore, computations showed that the use of the DMRG-TCCSD method is indispensable in order to determine the proper structure of the low lying energy spectrum in strongly correlated systems \cite{veis2018intricate}. 
In addition to these computational features, the DMRG-TCC approach is a promising candidate for a black-box quasi multi-reference scheme as big parts of the program already provide a black-box implementation up to a numerical threshold.
This makes it accessible to a broad class of researchers from various fields of study. 
For an alternative multi-reference CC method that makes use of matrix product states and a modified DMRG algorithm, we refer to the linearized CCSD theory of Sharma and Alavi~\cite{sharma2015multireference}.\newline
\indent
The mathematical analysis of CC schemes is far from being complete, especially with regard to multi-reference methods, however, many important steps have already been taken.
The list of fundamental and mathematical chemistry articles aiming to describe the existence and nature of solutions of CC equations is too long to be summarized here.
We will limit our discussion to a short selection of publications addressing such fundamental issues of chemistry.\newline
\indent
As a system of polynomial equations the CC equations can have real or, if the cluster operator is truncated, complex solutions.
A standard tools to compute a solution of these non-linear equations is the Newton--Raphson and the quasi Newton method.
However, these methods may diverge if the Jacobian or the approximated Jacobian become singular.
This is in particular the case when strongly correlated systems are considered.
These and other related aspects of the CC theory have been addressed by {\v{Z}}ivkovi{\'c} and Monkhorst \cite{vzivkovic1977existence,vzivkovic1978analytic} and Piecuch {\it et al.} \cite{piecuch1990coupled}.
Significant advances in the understanding of the nature of multiple solutions of single-reference CC have been made by Kowalski and Jankowski \cite{kowalski1998towards} and by Piecuch and Kowalski \cite{piecuch2000computational}.
An interesting attempt to address the existence of a cluster operator and cluster expansion in the open-shell case was done by Jeziorski and Paldus \cite{jeziorski1989valence}.
%
%Despite the valuable contributions in these works, and the fact that they are considered as fundamental and mathematical in the chemistry community, we highlight that these works are not entirely mathematically rigorous. 
%
The first advances within the rigorous realm of local functional analysis were performed by R.S. and Rohwedder, providing analyses of the closed-shell CC method for non multi-configuration systems \cite{schneider2009analysis,rohwedder2013continuous,rohwedder2013error}. 
Since then, the local mathematical analysis of CC schemes was extended by A.L. and S.K. analyzing the extended CC method \cite{laestadius2018analysis} and revisiting the bivariational approach \cite{lowdin1983stability,arponen1983variational}, and F.M.F. {\it et al.} providing the first local mathematical analysis of a multi-reference scheme, namely the CC method tailored by tensor network states (TNS-TCC) \cite{faulstich2018analysis}.
As this mathematical branch of CC theory is very young, channeling abstract mathematical results into the quantum-chemistry community is a highly interdisciplinary task merging both fields.
A first attempt in this direction was done by A.L. and F.M.F. linking the physical assumption of a HOMO-LUMO gap to the somewhat abstract G\aa rding inequality and in that context presenting key aspects of Refs.~\cite{schneider2009analysis,rohwedder2013continuous,rohwedder2013error,laestadius2018analysis} from a more quantum chemical point of view \cite{laestadius2018coupled}. \newline
\indent With this article, we aim to bridge %in the same spirit 
the mathematical results in Ref.~\cite{faulstich2018analysis} of the TNS-TCC method into the quantum-chemistry community.
Furthermore, we derive chemical properties of the TCC method and extend these results with a numerical study on the complete active space (CAS) dependence of the DMRG-TCCSD error.

\section*{II. The DMRG-TCC Method}
\label{Sec:DMRGMethod}
As a post-Hartree--Fock method, the TCC approach was introduced by Kinoshita {\it et al.} in Ref.~\cite{kinoshita2005coupled} as an alternative to the expensive and knotty conventional multi-reference methods.
It divides the cluster operator into a complete active space (CAS) part, denoted $\hat S$, and an external (ext) part $\hat T$, i.e., the wave function is parametrized as
\begin{equation*}
|\Psi_{\mathrm{TCC}}\rangle = \exp(\hat T)\exp(\hat S)|\Psi_{\mathrm{HF}}\rangle~.
\end{equation*}
Separating the cluster operator into several parts goes back to Piecuch {\it et al.} \cite{piecuch1993state, piecuch1994state}.
In this formulation the linked CC equations are given by
\begin{equation}
\label{eq:linkedTCC}
\left\lbrace
\begin{aligned}
E^{(\mathrm{TCC})} &=\langle\Psi_{\mathrm{HF}}|e^{-\hat S}e^{-\hat T} \hat H e^{\hat T}e^{\hat S }|\Psi_{\mathrm{HF}}\rangle~,\\
0&=\langle\Psi_{\mu}|e^{-\hat S }e^{-\hat T} \hat H e^{\hat T}e^{\hat S}|\Psi_{\mathrm{HF}}\rangle~.
\end{aligned}\right.
\end{equation}
Computing $|\Psi_{\mathrm{CAS}}\rangle=e^{\hat S}|\Psi_{\mathrm{HF}}\rangle$ first and keeping it fixed for the dynamical correction via the CCSD method restricts the above equations to $|\Psi_{\mu}\rangle$ not in the CAS, i.e., $\langle \Psi|\Psi_{\mu}\rangle = 0$ for all $|\Psi\rangle$ in the CAS (we say that $|\Psi_{\mu}\rangle$ is in the $L^2$-orthogonal complement of the CAS).
We emphasize that this includes mixed states, e.g., $|\Psi_{IJ}^{AB}\rangle$ where $|\Psi_{I}^{A}\rangle$ is an element of the CAS but $|\Psi_{J}^{B}\rangle$ is not.
We consider a CAS created by spin-orbitals $\mathscr{B}_{\mathrm{CAS}}=\{\chi_1,...,\chi_k\}$, forming a subspace of the full configuration interaction (FCI) space created by the entire set of spin-orbitals $\mathscr{B}=\{\chi_1,...,\chi_k,...,\chi_K\}$.
We here assume the spin-orbitals to be eigenfunctions of the system's Fock operator.
Note that the following analysis can be applied for any single-particle operator fulfilling the properties used in Ref.~\cite{faulstich2018analysis} -- not only the Fock operator.

Based on the single reference approach, the TCC method needs a large CAS to cover most of the static correlations.
As the size of the CAS scales exponentially, an efficient approximation scheme for strongly correlated systems is indispensable for the TCC method to have practical significance.
One of the most efficient schemes for static correlation is the DMRG method \cite{chan2011density}. 
Going back to the physicists S.~R. White and R.~L.~Martin \cite{white1999ab}, it was introduced to quantum chemistry as an alternative to the CI or CC approach.
However, the major disadvantage of the DMRG is that in order to compute dynamical correlation high bond dimensions (tensor ranks) may be necessary, making the DMRG a potentially costly method \cite{veis2018intricate,chan2011density}. 
Nevertheless, as a TNS-TCC method, the DMRG-TCC approach is an efficient method since the CAS is supposed to cover the statically correlated orbital of the system.
This avoids the DMRG method's weak point and allows to employ a much larger CAS compared to the traditional CI-TCC method.\newline
\indent
A notable advantage of the TCC approach over the conventional MRCC methods is that excitation operators commute.
This is due to the fact that the Hartee--Fock method yields a single reference solution $|\Psi_{\mathrm{HF}}\rangle$, which implies that separating the cluster operator corresponds to a partition of excitation operators.
Hence, $\hat S$ and $\hat T$ commute.
This makes the DMRG-TCC method's analysis much more accessible than conventional MRCC methods and therewith facilitates establishing sound mathematical results \cite{faulstich2018analysis}. 
Note also that the dynamical correlation via the CCSD method distinguishes the DMRG-TCC method from standard CAS-SCF methods, which suffer from an imbalance in the considered correlation making these methods inapplicable for chemical phenomena like surface hopping.  
Consequently, due to its more balanced correlation the DMRG-TCC approach is much better suited for these phenomena, which widens the horizon of possible applications. 
We remark, however, that the computationally most demanding step of the 
DMRG-TCC calculation is the DMRG part, and its cost  
increases rapidly with $k$.
Thus the above application requires further investigations.
Alternative to the dynamical correction via the CC approach, the {\it DMRG-MRCI} method in Ref.~\cite{saitow2013multireference} utilizes an {\it internally contracted CI} algorithm different from a conventional CI calculation.

\section*{III. Chemical Properties of The DMRG-TCC}
It is desired that quantum-chemical computations possess certain features representing the system's chemical and physical behavior. 
Despite their similarity, the CC and TCC method have essentially different properties, which are here elaborated. 
A basic property of the CC method is the equivalence of linked and unlinked CC equations.
We point out that this equivalence is in general not true for the DMRG-TCCSD scheme.
This is a consequence of the CAS ansatz since it yields mixed states, i.e., two particle excitations with one excitation into the CAS.
The respective overlap integrals in the unlinked CC equations will then not vanish unless the single excitation amplitudes are equal to zero. 
Generalizing this result for rank complete truncations of order $n$ we find that all excitation amplitudes need to be zero but for the $n$-th one. 
This is somewhat surprising as the equivalence of linked and unlinked CC equations holds for rank complete truncations of the single-reference CC method. \newline
\indent For the sake of simplicity we show this results for the DMRG-TCCSD method. The general case can be proven in similar a fashion.
We define the matrix representation $\mathbf{T}$ with elements $T_{\mu,\nu}=\langle\Psi_{\mu}|e^{\hat T}|\Psi_{\nu}\rangle$ for $\mu,\nu\notin \mathrm{CAS}$.
Note that, as $T$ increases the excitation rank, $\mathbf{T}$ is an atomic lower triangular matrix and therefore not singular. 
Assuming that the linked CC equations hold, the non-singularity of $\mathbf{T}$ yields
\begin{align*}
A_{\mu}
:&=
\sum_{\nu\notin \mathrm{CAS}} T_{\mu,\nu}\langle \Psi_{\nu}|e^{-{\hat T}}\hat He^{{\hat T}}|\Psi_\mathrm{HF}\rangle \\
&=
\sum_{\nu \notin \mathrm{CAS}} \langle\Psi_{\mu}|e^{\hat T}|\Psi_{\nu}\rangle \langle \Psi_{\nu}|e^{-{\hat T}}\hat He^{{\hat T}}|\Psi_{\mathrm{HF}}\rangle= 0~.\\
\end{align*}
As the full projection manifold is complete under de-excitation, we obtain that
\begin{align}
A_{\mu}
&=
\langle\Psi_{\mu}|\hat He^{{\hat T}}|\Psi_{\mathrm{HF}}\rangle
-E_{\mathrm{0}}\langle \Psi_{\mu}|e^{\hat T}|\Psi_{\mathrm{HF}}\rangle \nonumber \\
\label{eq:SubeqProofLinkVSUnLink}
&\quad-\sum_{\gamma\in \mathrm{CAS}}\langle\Psi_{\mu}|e^{\hat T}|\Psi_{\gamma}\rangle\langle\Psi_{\gamma}|\hat He^{{\hat T}}|\Psi_{\mathrm{HF}}\rangle~. 
\end{align}
Note that the first two terms on the r.h.s. in Eq.~\eqref{eq:SubeqProofLinkVSUnLink} describe the unlinked CC equations.
We analyze the last term on the r.h.s. in Eq.~\eqref{eq:SubeqProofLinkVSUnLink} by expanding the inner products, i.e.,
\begin{align*}
\langle\Psi_{\mu}|e^{\hat T}|\Psi_{\gamma}\rangle
&=
\langle\Psi_{\mu}|\Psi_{\gamma}\rangle+ \langle\Psi_{\mu}|{\hat T}|\Psi_{\gamma}\rangle\\
&\quad+\frac{1}{2}\langle\Psi_{\mu}|{\hat T}^2|\Psi_{\gamma}\rangle+... \quad.
\end{align*}
The first term in this expansion vanishes due to orthogonality.
The same holds true for all terms where $\hat T$ enters to the power of two or higher.
However, as the external space contains mixed states, we find that $\langle\Psi_{\mu}|\hat T|\Psi_{\gamma}\rangle$ is not necessarily zero, namely, for $\langle \Psi_{\mu}| =\langle \Psi_{\alpha}|\wedge \langle \Psi_{\beta}|$ and $|\Psi_{\gamma}\rangle = |\Psi_{\beta}\rangle$ with $\alpha\in\mathrm{ext}$ and $\beta\in\mathrm{CAS}$. This proves the claim. \newline
\indent Subsequently, we elaborate the size extensivity of the DMRG-TCCSD.
Let two DMRG-TCCSD wave functions for the individual subsystems $A$ and $B$ be
\begin{align*}
| \Psi_{\mathrm{DMRG-TCC}}^{(\mathrm{A})} \rangle
&=
\exp(\hat S_{\mathrm{A}})\exp(\hat T_{\mathrm{A}})| \Psi_{\mathrm{HF}}^{(\mathrm{A})} \rangle~,\\
| \Psi_{\mathrm{DMRG-TCC}}^{(\mathrm{B})} \rangle
&=
\exp(\hat S_{\mathrm{B}})\exp(\hat T_{\mathrm{B}})| \Psi_{\mathrm{HF}}^{(\mathrm{B})} \rangle~.
\end{align*}
The corresponding energies are given by
\begin{align*}
E_A =\langle \Psi_{\mathrm{HF}}^{(\mathrm{A})} |\hat{ \bar H}_A| \Psi_{\mathrm{HF}}^{(\mathrm{A})} \rangle~,\quad
E_B =\langle \Psi_{\mathrm{HF}}^{(\mathrm{B})} |\hat{ \bar H}_B| \Psi_{\mathrm{HF}}^{(\mathrm{B})} \rangle~,
\end{align*}
and the amplitudes fulfill
\begin{align*}
0 =\langle \Psi_{\mu}^{(\mathrm{A})} | \hat{\bar H}_A| \Psi_{\mathrm{HF}}^{(\mathrm{A})} \rangle~,\quad
0 =\langle \Psi_{\mu}^{(\mathrm{B})} | \hat{\bar H}_B| \Psi_{\mathrm{HF}}^{(\mathrm{B})} \rangle~,
\end{align*}
in terms of the effective, similarity-transformed Hamiltonians
\begin{align*}
\hat{\bar H}_A &= \exp(-\hat S_{\mathrm{A}}-\hat T_{\mathrm{A}}) \hat H_A \exp(\hat S_{\mathrm{A}}+\hat T_{\mathrm{A}})~,\\
\hat{\bar H}_B &= \exp(-\hat S_{\mathrm{B}}-\hat T_{\mathrm{B}}) \hat H_B \exp(\hat S_{\mathrm{B}}+\hat T_{\mathrm{B}})~.
\end{align*}
The Hamiltonian of the compound system of the non-interacting subsystems can be written as
$\hat H_{\mathrm{AB}} =\hat H_{\mathrm{A}}+\hat H_{\mathrm{B}}$.
Since the TCC approach corresponds to a partitioning of the cluster amplitude we note that $\hat{\bar H}_{\mathrm{AB}} = \hat{ \bar H}_{\mathrm{A}} + \hat{\bar H}_{\mathrm{B}}$
for 
\begin{align*}
\hat{\bar H}_{\mathrm{AB}}
=&
\exp(-\hat S_{\mathrm{A}}-\hat S_{\mathrm{B}}-\hat T_{\mathrm{A}}-\hat T_{\mathrm{B}}) \\
&\times \hat H_{\mathrm{AB}}
\exp(\hat S_{\mathrm{A}}+\hat T_{\mathrm{B}}+\hat T_{\mathrm{A}}+\hat T_{\mathrm{B}})~.
\end{align*}
With $
| \Psi_{\mathrm{HF}}^{(\mathrm{AB})}  \rangle= | \Psi_{\mathrm{HF}}^{(\mathrm{A})} \rangle  \wedge | \Psi_{\mathrm{HF}}^{(\mathrm{B})}  \rangle$, the energy of the compound systems can be written as
\begin{align*}
E_{\mathrm{AB}}
&=
\langle \Psi_{\mathrm{HF}}^{(\mathrm{AB})}| \hat{\bar H}_{\mathrm{AB}} | \Psi_{\mathrm{HF}}^{(\mathrm{AB})}\rangle\\
&=
\big(\langle \Psi_{\mathrm{HF}}^{(\mathrm{A})}|  \wedge \langle \Psi_{\mathrm{HF}}^{(\mathrm{B})} |\big)\big( \hat{\bar H}_{\mathrm{A}} + \hat{\bar H}_{\mathrm{B}}\big)\big( | \Psi_{\mathrm{HF}}^{(\mathrm{A})} \rangle  \wedge | \Psi_{\mathrm{HF}}^{(\mathrm{B})}  \rangle\big)\\
&=
\langle \Psi_{\mathrm{HF}}^{(\mathrm{A})}|\hat{\bar H}_{\mathrm{A}}| \Psi_{\mathrm{HF}}^{(\mathrm{A})} \rangle
+
\langle \Psi_{\mathrm{HF}}^{(\mathrm{B})}|\hat{\bar H}_{\mathrm{B}}| \Psi_{\mathrm{HF}}^{(\mathrm{B})} \rangle
=
E_{\mathrm{A}}+E_{\mathrm{B}}~.
\end{align*} 
It remains to show that 
\begin{align*}
| \Psi_{\mathrm{DMRG-TCC}}^{(\mathrm{AB})} \rangle
=
\exp(\hat S_{\mathrm{A}}+\hat S_{\mathrm{B}}+\hat T_{\mathrm{A}}+\hat T_{\mathrm{B}})| \Psi_{\mathrm{HF}}^{(\mathrm{AB})}  \rangle
\end{align*}
solves the Schr\"odinger equation, i.e., for all $\langle \Psi_{\mu}^{(\mathrm{AB})} |$ holds $\langle \Psi_{\mu}^{(\mathrm{AB})} | \hat{\bar H}_{\mathrm{AB}}| \Psi_{\mathrm{HF}}^{(\mathrm{AB})}  \rangle = 0$.
Splitting the argument into three case, we note that
\begin{align*}
\langle \Psi_{\mu}^{(\mathrm{A})} \Psi_{(\mathrm{HF})}^{(\mathrm{B})} |  \hat{\bar H}_{\mathrm{AB}}| \Psi_{\mathrm{HF}}^{(\mathrm{AB})}  \rangle
&=\langle \Psi_{\mu}^{(\mathrm{A})} | \hat{\bar H}_{\mathrm{A}}| \Psi_{\mathrm{HF}}^{(\mathrm{A})}  \rangle
=
0~,
\\
\langle \Psi_{(\mathrm{HF})}^{(\mathrm{A})} \Psi_{\mu}^{(\mathrm{B})} | \hat{\bar H}_{\mathrm{AB}}| \Psi_{\mathrm{HF}}^{(\mathrm{AB})}  \rangle
&= \langle \Psi_{\mu}^{(\mathrm{B})} |\hat{\bar H}_{\mathrm{B}}| \Psi_{\mathrm{HF}}^{(\mathrm{B})}  \rangle
=
0~,
\\
\langle \Psi_{\mu}^{(\mathrm{A})} \Psi_{\mu}^{(\mathrm{B})} | \hat{\bar H}_{\mathrm{AB}}| \Psi_{\mathrm{HF}}^{(\mathrm{AB})}  \rangle
&=0~,
\end{align*}
where $\langle \Psi^{(\mathrm{A})} \Psi^{(\mathrm{B})}| = \langle \Psi^{(\mathrm{A})}|\wedge \langle\Psi^{(\mathrm{B})}|$.
This proves the energy size extensivity for the untruncated TCC method. 
From this we conclude the energy size extensivity for the DMRG-TCCSD scheme, because the truncation only affects the product states $\langle \Psi_{\mu}^{(\mathrm{A})} \Psi_{\mu}^{(\mathrm{B})} |$ and these are zero in the above projection.

Looking at TCC energy expression we observe that due to the Slater--Condon rules, these equations are independent of CAS excitations higher than order three, i.e., amplitudes of $\hat S_n$ for $n>3$.
More precisely, due to the fact that in the TCCSD case external space amplitudes can at most contain one virtual orbital in the CAS, the TCCSD amplitude expressions become independent of $\hat S_4$, i.e.,  
\begin{align*}
\langle \phi_{ij}^{a'a}|\hat H | \psi_{klmn}^{b'c'd'e'}\rangle
=0~,
\end{align*}
where the primed variables $a',b',c',d',e'$ describe orbitals in the CAS, the non-primed variable $a$ describes an orbital in the external part and $i,j,k,l,m,n$ are occupied orbitals.  
Note, this does not imply that we can restrict the CAS computation to a manifold characterizing excitations with rank less or equal to three as for strongly correlated systems these can still be relevant.
However, it reduces the number of terms entering the DMRG-TCCSD energy computations significantly. \newline
\indent In contrast to the originally introduced CI-TCCSD method taking only $\hat S_n$ for $n=1,2$ into account \cite{kinoshita2005coupled}, the additional consideration of $\hat S_3$ corresponds to an exact treatment of the CAS contributions to the energy.
We emphasize that the additional terms do not change the methods complexity.
This is due to the fact that including the CAS triple excitation amplitudes will not exceed the dominating complexities of the CCSD \cite{myhre2016multilevel} nor of the DMRG.
However, the extraction of the CI-triples from the DMRG wave function is costly and a corresponding efficiency investigation is left for future work.
%
%As a last remark, the number of excitation amplitudes, denoted $\sharp \mathcal{J}_{ext}$, for an $N$-electron problem with a given separation of the basis set $\mathscr{B}$ at the $k$-th spin orbital is given by 
%\begin{equation*}
%%\begin{aligned}
%\sharp\mathcal{J}_{ext}
%=
%\sum_{d=1}^2\binom{N}{d}\binom{K-k}{k} \in \mathcal{O}(N^2(K-k)^2).
%&=
%N(K-k)\left(1+\frac{1}{2}(N-1)(K-k-1)\right)~.
%\end{aligned}
%\end{equation*}

\section*{IV. Analysis of the DMRG-TCC}
\label{sec:AnalysisDMRGTCC}
In the sequel we discuss and elaborate mathematical properties of the TCC approach and their influence on the DMRG-TCC method. 
The presentation here is held brief and the interested reader is referred to Ref.~\cite{faulstich2018analysis} and the references therein for further mathematical details.

\subsection*{A. The Complete Active Space Choice}
\label{SubSec:ThecompleteActiveSpaceChoice}
As pointed out in the previous section, the TCC method relies on a {\it well-chosen} CAS, i.e., a large enough CAS that covers the system's static correlation.
Consequently, we require a quantitative measurement for the quality of the CAS, which presents the first obstacle for creating a non-empirical model 
since the chemical concept of correlation is not well-defined \cite{lyakh2011multireference}.  
In the DMRG-TCC method, we use a quantum information theory approach to classify the spin-orbital correlation.
This classification is based on the {\it mutual-information} 
\begin{align*}
I_{i|j} = S(\rho_{\{i\}})+S(\rho_{\{j\}})-S(\rho_{\{i,j\}})~.
\end{align*}
This two particle entropy is defined via the {\it von Neumann entropy} $S(\rho) = - \mathrm{Tr}(\rho\ln\rho )$ of the reduced density operators $\rho_{\{X\}}$ \cite{szalay2017correlation}. 
Note that the mutual-information describes two-particle correlations, for a more general connection between multi-particle correlations and $\xi$-correlations we refer the reader to Szalay {\it et.~al }\cite{szalay2017correlation}. 
We emphasize that in practice this is a basis dependent quantity, which is in agreement with the chemical definition of correlation concepts \cite{lyakh2011multireference}.
We identify pairs of spin-orbitals contributing to a high mutual information value as strongly correlated, the pairs contributing to the plateau region as non-dynamically correlated and the pairs contributing to the mutual information tail as dynamically correlated 
(see Fig.~\ref{fig:mutualinfo-fullcas}). 
The mutual-information profile can be well approximated from a prior DMRG computation on the full system.
Due to the size of the full system we only compute a DMRG solution of low {\it bond dimension} (also called {\it tensor rank}).
These low-accuracy calculations, however, already provide a good qualitative
entropy profile, i.e., the shapes of profiles obtained for low bond dimension, 
$M$, agree well with the the ones obtained in the FCI limit. 
Here, we refer to Fig.~\ref{fig:entropy-fullcas} and Fig.~\ref{fig:mutualinfo-fullcas} showing the single orbital entropy and mutual information profiles, respectively, for various $M$ values and for three different geometries of the N$_2$ molecule. 
The orbitals with large entropies can be identified from the low-$M$ calculations providing a black-box tool to form the CAS space including the strongly correlated orbitals.

A central observation is that for $\mathscr{B}_{\mathrm{CAS}}=\{\chi_1,...,\chi_N\}$ (i.e. $k=N$), the DMRG-TCCSD becomes the CCSD method and for $\mathscr{B}_{\mathrm{CAS}}=\{\chi_1,...,\chi_K\}$ (i.e. $k=K$), it is the DMRG method.
We recall that the CCSD method can not resolve static correlation and the DMRG method needs high tensor ranks for dynamically correlated systems.
This suggests that the error obtains a minimum for some $k$ with $N\leq k\leq K$, i.e., there exists an optimal choice of $k$ determining the basis splitting and therewith the choice of the CAS.
Note that this feature becomes important for large systems since high bond dimensions become simply impossible to compute with available methods. 

\subsection*{B. Local Analysis of the DMRG-TCC Method}
The CC method can be formulated as {\it non-linear Galerkin scheme} \cite{schneider2009analysis}, which is a well-established framework in numerical analysis to convert the continuous Schr\"odinger equation to a discrete problem.
For the DMRG-TCC method a first local analysis was performed in Ref.~\cite{faulstich2018analysis}.
There, a quantitative error estimate with respect to the basis truncation was established.
F.M.F. {\it et al.} showed under certain assumptions (Assumption A and B in the sequel) that the DMRG-TCC method possesses a locally unique and {\it quasi-optimal} solution (cf. Sec.~4.1 in Ref.~\cite{faulstich2018analysis}).
%
%The concept of quasi optimality goes back to C\'ea \cite{cea1964approximation}, who originally introduced it for linear Galerkin schemes.
%
In case of the DMRG-TCC method the latter means:
For a fixed basis set the CC solution tailored by a DMRG solution on a fixed CAS is up to a multiplicative constant the best possible solution in the approximation space defined by the basis set.     
In other words, the CC method provides the best possible dynamical correction for a given CAS solution such as a DMRG solution.\newline
\indent 
Note that local uniqueness ensures that for a fixed basis set, the computed DMRG-TCC solution is unique in a neighborhood around the exact solution.
We emphasize that this result is derived under the assumption that the CAS solution is fixed.
Consequently, for different CAS solutions we obtain in general different TCC solutions, i.e., different cluster amplitudes.\newline
\indent
Subsequently, parts of the results in Ref.~\cite{faulstich2018analysis} are explained in a setting adapted to the theoretical chemistry perspective.
The TCC function is given by 
%\begin{equation*}
$f(t;s)=\langle  \Psi_{\mu}| e^{-\hat S}e^{-\hat T}\hat H e^{\hat T}e^{\hat S}|\Psi_{\mathrm{HF}} \rangle, %\quad \Psi_\mu\nsubseteq \mathrm{CAS},
$
%\end{equation*}
for $|\Psi_\mu\rangle$ not in the CAS. 
Note that we use the convention where small letters $s,t$ correspond to cluster amplitudes, whereas capital letters $\hat S,\, \hat T$ describe cluster operators. 
The corresponding TCC energy expression is given by 
\begin{equation*}
\mathcal E (t;s) = \langle \Psi_{\mathrm{HF}}|  e^{-\hat S}e^{-\hat T}\hat He^{\hat T}e^{\hat S}|\Psi_{\mathrm{HF}}  \rangle~.
\end{equation*}
Consequently, the linked TCC equations \eqref{eq:linkedTCC} then become
\begin{equation*}
\left\lbrace
\begin{aligned}
E^{(\mathrm{TCC})} &= \mathcal E(t;s)~,\\ %&=\left\langle\phi_0,e^{-T^{\mathrm{CAS}}}e^{-T^{\mathrm{ext}}}He^{T^{\mathrm{CAS}}}e^{T^{\mathrm{ext}}}\phi_0 \right\rangle~,\\
0&=f(t;s) ~.\\
\end{aligned}
\right.
\end{equation*}
Within this framework the locally unique and quasi-optimal solutions of the TCC method were obtained under two assumptions (see Assumption A and B in Ref.~\cite{faulstich2018analysis}).\newline 
\indent
First, Assumption A requires that the Fock operator $\hat F$ is {\it bounded} and satisfies a so-called {\it G\aa rding inequality}. 
For a more detailed description of these properties in this context we refer the reader to Ref.~\cite{laestadius2018coupled}.  
Second, it is assumed that there exists a CAS-ext gap in the spectrum of the Fock operator, i.e., there is a gap between the $k$-th and the $k + 1$-st orbital energies.
Note, that spectral gap assumptions (cf. HOMO-LUMO gap) are standard in the analysis of dynamically correlated systems.
To be applicable to multi-configuration systems, the analysis in Ref.~\cite{faulstich2018analysis} rests on a CAS chosen such that the $k$-th and the $(k+1)$-st spin orbital are non-degenerate. 
Intuitively, this gap assumption means that the CAS captures the static correlation of the system.\newline
\indent
Assumption B is concerned with the fluctuation operator $\hat W = \hat H - \hat F$. 
This operator describes the degeneracy of the system since it is the difference of the Hamiltonian and a single particle operator.
Using the similarity transformed $\hat W$ and fixing the CAS amplitudes $s$, the map 
\begin{align*}
t\mapsto e^{-\hat T}e^{-\hat S}\hat W e^{\hat S}e^{\hat T}|\Psi_{\mathrm{HF}}\rangle
\end{align*}
is assumed to have a {\it small enough} Lipschitz-continuity constant (see Eq.~(20) in Ref.~\cite{faulstich2018analysis}).
The physical interpretation of this Lipschitz condition is at the moment unclear.
\subsection*{C. Error Estimates for the DMRG-TCC Method}
A major difference between the CI and CC method is that the CC formalism is not variational.
Hence, it is not evident that the CC energy error decays quadratically with respect to the error of the wave function or cluster amplitudes.
Note that the TCC approach represents merely a partition of the cluster operator, however, its error analysis is more delicate than the traditional CC method's analysis.   
The TCC-energy error is measured as difference to the FCI energy. 
Let $|\Psi^*\rangle$ describe the FCI solution on the whole space, i.e., $\hat H |\Psi^*\rangle=E|\Psi^*\rangle$. 
Using the exponential parametrization and the above introduced separation of the cluster operator, we have 
\begin{equation}
\label{eq:PsiStartExponentialParam}
|\Psi^*\rangle
=\exp(\hat T^*)\exp(\hat S^*)|\Psi_{\mathrm{HF}}\rangle~.
\end{equation}
An important observation is that the TCC approach ignores the coupling from the external space into the CAS.
It follows that the FCI solution on the CAS $|\Psi_{\mathrm{FCI}}^{\mathrm{CAS}}\rangle=\exp(\hat S_{\mathrm{FCI}})|\Psi_{\mathrm{HF}}\rangle$ is an approximation to the projection of $|\Psi^*\rangle$ onto the CAS
\begin{align*}
| \Psi_{\mathrm{FCI}}^{\mathrm{CAS}}\rangle
\approx
\hat P|\Psi^*\rangle = \exp(\hat S^*)|\Psi_{\mathrm{HF}}\rangle~,
\end{align*}
where $\hat P=|\Psi_{\mathrm{HF}}\rangle\langle\Psi_{\mathrm{HF}}|+\sum_{\mu\in \mathrm{CAS}}|\Psi_{\mu}\rangle\langle\Psi_{\mu}|$ is the $L^2$-orthogonal projection onto the CAS.
For a reasonably sized CAS the FCI solution $|\Psi_{\mathrm{FCI}}^{\mathrm{CAS}}\rangle$ is rarely computationally accessible and we introduce the DMRG solution on the CAS as an approximation of $|\Psi_{\mathrm{FCI}}^{\mathrm{CAS}}\rangle$,
\begin{align*}
|\Psi_{\mathrm{DMRG}}^{\mathrm{CAS}}\rangle=\exp(\hat S_{\mathrm{DMRG}})|\Psi_{\mathrm{HF}}\rangle\approx|\Psi_{\mathrm{FCI}}^{\mathrm{CAS}}\rangle ~.
\end{align*}
Tailoring the CC method with these different CAS solutions leads in general to different TCC solutions. 
In the case of $|\Psi_{\mathrm{FCI}}^{\mathrm{CAS}}\rangle$, the TCC method yields the best possible solution with respect to the chosen CAS, i.e., $f(t_{\mathrm{CC}}^*;s_{\mathrm{FCI}})=0$.
This solution is in general different from $t_{\mathrm{CC}}$ fulfilling $f(t_{\mathrm{CC}};s_{\mathrm{DMRG}})=0$ and its truncated version $t_{\mathrm{CCSD}}$ satisfying $P_{\mathrm{Gal}}f(t_{\mathrm{CCSD}};s_{\mathrm{DMRG}})=0$, where $P_{\mathrm{Gal}}$ denotes the $l^2$-orthogonal projection onto the corresponding Galerkin space.
In the context of the DMRG-TCC theory, the Galerkin space represents a truncation in the excitation rank of the cluster operator, e.g., DMRG-TCCD, DMRG-TCCSD, etc.\newline
\indent
For the  following argument, suppose that an appropriate CAS has been fixed.  
The total DMRG-TCC energy error $\Delta E$ can be estimated as \cite{faulstich2018analysis}
\begin{equation}
\label{eq:EestMain}
\begin{aligned}
\Delta E &=
| \mathcal E(t_{\mathrm{CCSD}};s_{\mathrm{DMRG}}) -\mathcal E(t^*;s^*)| \\
&\leq
\Delta \varepsilon + \Delta \varepsilon_{\mathrm{CAS}} + \Delta \varepsilon_{\mathrm{CAS}}^*~,
\end{aligned}
\end{equation}
where each term of the r.h.s. in Eq.~\eqref{eq:EestMain} is now discussed.
As a technical remark, the norms on either the Hilbert space of cluster amplitudes or wave functions are here simply denoted $\Vert \cdot\Vert$. 
These norms are not just the $l^2$- or $L^2$-norm, respectively, but also measure the kinetic energy. 
It should be clear from context which Hilbert space is in question and we refer to Ref.~\cite{rohwedder2013continuous} for formal definitions.
The first term is defined as 
\begin{align*}
\Delta\varepsilon = | \mathcal E(t_{\mathrm{CCSD}};s_{\mathrm{DMRG}}) 
- \mathcal{E}(t_{\mathrm{CC}};s_{\mathrm{DMRG}})|~,
\end{align*} 
which describes the truncation error of the CCSD method tailored by $|\Psi_{\mathrm{DMRG}}^{\mathrm{CAS}}\rangle$.
We emphasize that the dynamical corrections via the CCSD and the untruncated CC method are here tailored by the same CAS solution. 
Hence, the energy error $\Delta\varepsilon$ corresponds to a single reference CC energy error, which suggests an analysis similar to Refs.~\cite{schneider2009analysis,rohwedder2013error}.
Indeed, the {\it Aubin--Nitsche duality method} \cite{aubin1967behavior,nitsche1968kriterium,ruchovec1969study} yields a quadratic \textit{a priori} error estimate in $\Vert t_{\mathrm{CCSD}} - t_{\mathrm{CC}} \Vert $ (and in terms of the Lagrange mulitpliers, see Theorem 29 in Ref.~\cite{faulstich2018analysis}). 
\newline
\indent 
Second, we discuss the term 
\begin{align*}
\Delta\varepsilon_{\mathrm{CAS}}
=
| \mathcal E(t_{\mathrm{CC}};s_{\mathrm{DMRG}}) - \mathcal{E}(t_{\mathrm{CC}};s_{\mathrm{FCI}}  )|~.
\end{align*}
Here, different CAS solutions with fixed external solutions are used to compute the energies.
This suggests that $\Delta\varepsilon_{\mathrm{CAS}}$ is connected with the error 
\begin{equation}
\begin{aligned}
\Delta E_{\mathrm{DMRG}}&=|\langle\Psi_{\mathrm{HF}}|e^{-\hat S_{\mathrm{DMRG}}} \hat P \hat H \hat Pe^{\hat S_{\mathrm{DMRG}}}\\
&\quad -e^{-\hat S_{\mathrm{FCI}}} \hat P \hat H \hat Pe^{\hat S_{\mathrm{FCI}}}|\Psi_{\mathrm{HF}}\rangle|~,
\end{aligned}
\label{eq:Ecas}
\end{equation}
describing the approximation error of the DMRG solution on the CAS (see Lemma 27 in Ref.~\cite{faulstich2018analysis}).
Indeed, 
\begin{equation}
\begin{aligned}
\label{eq:SecondErrorTerm}
\Delta \varepsilon_{\mathrm{CAS}} 
&\lesssim  
\Delta E_{\mathrm{DMRG}} + \Vert t_{\mathrm{CC}} - t_{\mathrm{CC}}^*\Vert^2 \\
&\quad+ \Vert ( \hat S_{\mathrm{DMRG}} 
- \hat S_{\mathrm{FCI}})|\Psi_{\mathrm{HF}}\rangle\Vert^2\\
&\quad+ \sum_{|\mu|=1} \varepsilon_\mu (t_{\mathrm{CC}}^*)_\mu^2~,
\end{aligned}
\end{equation}
with $\varepsilon_{\mu}=\varepsilon_{I_1...I_n}^{A_1...A_n}= \sum_{j=1}^n (\lambda_{A_j} - \lambda_{I_j})$, for $1\leq n\leq k$, where $\lambda_i$ are the orbital energies.
The $\varepsilon_{\mu}$ are the (translated) Fock energies, more precisely, $\hat F |\Psi_{\mu}\rangle = (\Lambda_0+\varepsilon_{\mu})|\Psi_{\mu}\rangle$, with $\Lambda_0 = \sum_{i=1}^{N}\lambda_i$.
Note that the wave function $|\Psi_{\mathrm{FCI}}^{\mathrm{CAS}}\rangle$ is in general not an eigenfunction of $\hat H$, however, it is an eigenfunction of the projected Hamiltonian $\hat P\hat H \hat P$. 
Eq.~\eqref{eq:Ecas}  involves the exponential parametrization. 
This can be estimated by the energy error of the DMRG wave function, denoted $\Delta \mathcal E_{\mathrm{DMRG}}$, namely,
\begin{equation}
\begin{aligned}
\label{eq:WavefunctionError}
\Delta E_{\mathrm{DMRG}}
&\leq
2\Delta \mathcal E_{\mathrm{DMRG}}\\
&\quad +\Vert \hat H \Vert ~ \Vert~|\Psi_{\mathrm{DMRG}}^{\mathrm{CAS}}\rangle-|\Psi_{\mathrm{FCI}}^{\mathrm{CAS}}\rangle~\Vert_{L^2}~.
\end{aligned} 
\end{equation}
In Sec.~V the energy error of the DMRG wave function is controlled by the threshold value $\delta \varepsilon_\mathrm{Tr}$, i.e., $ \Delta \mathcal E_{\mathrm{DMRG}}(\delta \varepsilon_\mathrm{Tr})$. 
Hence, for well chosen CAS the difference $\Vert~|\Psi_{\mathrm{DMRG}}^{\mathrm{CAS}}\rangle-|\Psi_{\mathrm{FCI}}^{\mathrm{CAS}}\rangle~\Vert_{L^2}$ is sufficiently small such that $\Delta E_{\mathrm{DMRG}}
\lesssim
2\Delta \mathcal E_{\mathrm{DMRG}}$ holds.
This again shows the importance of a well-chosen CAS.
Furthermore, the last term in Eq.~\eqref{eq:SecondErrorTerm} can be eliminated via orbital rotations, as it is a sum of single excitation amplitudes.\newline
\indent 
Finally, we consider 
\begin{align}
\label{eq:MethodologicalError}
\Delta \varepsilon^*_{\mathrm{CAS}}
=
| \mathcal E(t_{\mathrm{CC}};s_{\mathrm{FCI}}) - \mathcal{E}(t^*;s^*  )| ~.
\end{align}
Since $(t^*,s^*)$ is a stationary point of $\mathcal E$ we have $D\mathcal E(t^*;s^*)=0$.
A calculation involving Taylor expanding $\mathcal E$ around $(t^*,s^*)$ (see Lemma 26 in Ref.~\cite{faulstich2018analysis})  yields
\begin{align}
\label{eq:MethodologicalBound}
\Delta \varepsilon_{\mathrm{CAS}}^*
\lesssim
\Vert   t_{\mathrm{CC}}- t^*\Vert^2+ \Vert s_{\mathrm{FCI}}-s^* \Vert_{l^2}^2~.
\end{align}
Note that the above error is caused by the assumed basis splitting, namely, the correlation from the external part into the CAS is ignored. 
Therefore, the best possible solution for a given basis splitting $(t_\mathrm{CC}^*,s_{\mathrm{FCI}})$ differs in general from the FCI solution $(t^*,s^*)$.  \newline
\indent
Combining now the three quadratic bounds gives an overall quadratic \textit{a priori} energy error estimate for the DMRG-TCC method.
The interested reader is referred to Ref.~\cite{faulstich2018analysis} for a more detailed treatment of the above analysis.
\subsection*{D. On the $k$-dependency of the Error Estimates}
\label{sec:onkdep}
The error estimate outlined above is for a fixed CAS, i.e., a particular basis splitting, and bounds the energy error in terms of truncated amplitudes. 
Because the TCC solution depends strongly on the choice of the CAS, it is motivated to further investigate the $k$-dependence of the error $\Delta E$. 
However, the above derived error bound has a highly complicated $k$ dependence since not only the amplitudes but also the implicit constants (in $\lesssim$) and norms depend on $k$. 
Therefore, the analysis in Ref.~\cite{faulstich2018analysis} is not directly applicable to take the full $k$ dependence into account. \newline % since the used Aubin-Nitsche trick exploits the existence of a CAS-ext gap.
\indent In the limit where $s_{\mathrm{DMRG}}\to s_{\mathrm{FCI}}$ we obtain that $t_{\mathrm{CC}}\to t_{\mathrm{CC}}^*$ since the TCC method is numerically stable, i.e., a small perturbation in $s$ corresponds to a small perturbation in the solution $t$. 
Furthermore, if we assume that $t_{\mathrm{CCSD}}\approx t_{\mathrm{CC}}$, which is reasonable for the equilibrium bond length of N$_2$, 
the error can be bound as
\begin{align}
\Delta E_k
&\leq 
C_k\Big(\sum_{|\mu|=1}(t_{\mathrm{CCSD}})_{\mu}^2
+
\left\Vert 
\begin{pmatrix}
t_{\mathrm{CCSD}} \\
s_{\mathrm{DMRG}}
\end{pmatrix}
-
\begin{pmatrix}
t^*\\
s^*
\end{pmatrix}
\right\Vert_{l^2}^2\Big)~.
\label{eq:error-k}
\end{align}
Here the subscript $k$ on $\Delta E_k$ and $C_k$ highlights the $k$-dependence. 
We remark that we here used the less accurate $l^2$ structure on the amplitude space compared to the $H^1$ structure in Eq.~\eqref{eq:MethodologicalBound}. 
This yields $k$-independent vectors
\[
\begin{pmatrix}
t_{\mathrm{CCSD}} \\
s_{\mathrm{DMRG}}
\end{pmatrix},\quad
\begin{pmatrix}
t^*\\
s^*
\end{pmatrix},
\]
as well as an $k$-independent $l^2$ norm. 
The $k$-depenence of $C_k$ will be investigated numerically
in more detail in Sec.~V\,B\,5.

\section*{V. The Splitting Error for N$_2$}
\label{Sec:TheSplittingError}
Including the $k$ dependence in the above performed error analysis explicitly 
is a highly non-trivial task involving many mathematical obstacles 
and is part of our current research.
Therefore, we here extend the mathematical results from Sec.~IV with a numerical investigation on this $k$ dependence.
Our study is presented for the N$_2$ molecule using the cc-pVDZ basis, which 
is a common basis for benchmark computations developed by Dunning and coworkers \cite{dunning1989gaussian}.
We investigate three different geometries of the Nitrogen dimer by stretching the molecule, thus the performance of DMRG-TCCSD method is assessed against DMRG and single reference CC methods for bond lengths $r=2.118\,a_0, 2.700\,a_0$, and $3.600\,a_0$. 
In the equilibrium geometry the system is weakly correlated implying that single reference CC methods yield reliable results.
For increasing bond length $r$ the system shows multi-reference character, i.e., static correlations become more dominant. 
For $r>3.5\,a_0$ this results in the variational breakdown of single reference CC methods \cite{kowalski2000renormalized}.
This breakdown can be overcome with the DMRG-TCCSD method once a large and well chosen CAS is formed, 
we therefore refer to the DMRG-TCCSD method as numerically stable with respect to the bond length
along the potential energy surface (PES).

As mentioned before, the DMRG method is in general less efficient to recover dynamic correlations since it 
requires large computational resources.
However, due to the specific CAS choice the computational resource for the DMRG part of the TCC scheme 
is expected to be significantly lower than a pure DMRG calculation for the same level of accuracy.

\subsection*{A. Computational Details}
\label{sec:numproc}

In practice, a routine application of the TCC method to strongly correlated molecular systems, i.e., 
to multi-reference problems, became possible only recently since it requires 
a very accurate solution in a large CAS including all static correlations. 
Tensor network state methods fulfill such a high accuracy criterion, but the efficiency of the 
TNS-TCCSD method strongly depends on various parameters of the involved algorithms. 
Some of these are defined rigorously while others are more heuristic from the mathematical point of view.
In this section we present the optimization steps for the most important parameters of the DMRG-TCCSD method and outline how the numerical error study in Sec.~V\,B is performed.

As elaborated in Sec.~II and IV\,A, the CAS choice is essential for the computational success of TNS-TCC methods. 
In addition, the error of the TNS method used to approximate the CAS part depends on various approximations. 
These include the proper choice of a finite dimensional basis to describe the chemical compound, the tensor network structure, and the mapping of the molecular orbitals onto the given network \cite{Szalay-2015}.  
Fortunately, all these can be optimized by utilizing concepts of quantum 
information theory, introduced in Sec.~IV\,A (see also the included references).
In the following, we restrict the numerical study to the DMRG-TCCSD method 
but the results presented here should also hold for other TNS 
approaches \cite{Murg-2010a,Nakatani-2013,Murg-2015,Szalay-2015,Gunst-2018}.\newline
\indent
In the DMRG-TCCSD case the tensor network topology in the CAS corresponds to a single branched tensor tree, i.e., a one dimensional topology. 
Thus permutations of orbitals along such an artificial chain effect the 
convergence for a given CAS choice \cite{Chan-2002a,Legeza-2003b}.
This orbital-ordering optimization can be carried out based on spectral graph 
theory \cite{Barcza-2011,Fertitta-2014} by minimizing the entanglement 
distance \cite{Rissler-2006}, defined as 
$I_{\rm dist} = \sum_{ij} I_{i|j} |i-j|^2$. 
In order to speed up the convergence of the DMRG procedure the configuration 
interaction based dynamically extended active space (CI-DEAS) method
is applied \cite{Legeza-2003b,Szalay-2015}.
In the course of these optimization steps, the single orbital entropy 
($S_i=S(\rho_{\{i\}})$) and the two-orbital mutual information $(I_{i|j})$ are calculated 
iteratively until convergence is reached. 
The size of the active space is systematically increased by including 
orbitals with the largest single site entropy values, which at the same 
time correspond to orbitals contributing to the largest matrix elements 
of the mutual information. 
Thus, the decreasingly ordered values of $S_i$ define the so-called 
CAS vector, CAS$_{\rm vec}$, which provides a guide in what order to 
extend the CAS by including additional orbitals. 
The bond dimensions $M$ (tensor rank) in the DMRG method can be kept fixed 
or adapted dynamically (Dynamic Block State Selection (DBSS) approach) 
in order to fulfill an {\it a priori} defined error 
margin \cite{Legeza-2003a,Legeza-2004b}.
Accurate extrapolation to the truncation free limit is possible as a 
function of the truncation error $\delta\varepsilon_{\rm Tr}$ \cite{Legeza-1996,Legeza-2003a}.\newline
\indent In our DMRG implementation~\cite{QCDMRG-Budapest} we use a spatial orbital basis, i.e., the local tensor space of a single orbital is 
$d=4$ dimensional.  
In this $\mathbb{C}^4$ representation an orbital can be empty, singly occupied with either a spin up or spin down electron, or doubly occupied with opposite spins.
Note, in contrast to Sec.~IV we need $N/2$ spatial orbitals to describe an $N$-electron wave function and similar changes apply to the size of the basis set so that we use $K\equiv K/2$ from here on. 	
The single orbital entropy therefore varies between 0 and $\ln d = \ln 4$, 
while the two-orbital mutual information varies between 0 
and $\ln d^2 = \ln 16$.\newline
\indent Next we provide a short description how to perform DMRG-TCCSD calculations in practice.
Note that we leave the discussion on the optimal choice of $k$ for the following sections.\newline
\indent $1)$ First the CAS is formed from the full orbital space by setting $k=K$.
DMRG calculations are performed iteratively with fixed low bond dimension (or with a large error margin) in order to determine the optimal ordering and the CAS-vector as described above.
Thus, the corresponding single-orbital entropy and mutual information are also calculated. 
These calculations already provide a good qualitative description of the 
entropy profiles with respect to the exact solution, i.e., strongly 
correlated orbitals can be identified.\newline
\indent $2)$ Using a given $N/2< k< K$ we form the CAS from the Hartree--Fock orbitals and the first $k-N/2$ virtual orbitals from the CAS$_{\rm vec}$, i.e., orbitals with the largest single orbital entropy values. 
We emphasize that these orbitals contribute to the largest matrix elements 
in $I_{i|j}$.
We carry out the orbital ordering optimization on the given CAS and perform 
a large-scale DMRG calculation with a low error threshold margin in order 
to get an accurate approximation of the $|\Psi^{\rm CAS}_{\rm FCI}\rangle$.
Note, that the DMRG method yields a normalized wave function, i.e., 
the overlap with the reference determinant $|\Psi_{\rm HF}\rangle$ is not
necessarily equal to one. \newline
\indent $3)$ Using the matrix product state representation of 
$|\Psi^{\rm FCI}_{\rm DMRG}\rangle$ obtained by the DMRG method we 
determine the 
zero reference overlap, single and double CI coefficients of 
the full tensor representation of the wave function. 
Next, these are used to calculate the $\hat S_1$ and $\hat S_2$ amplitudes, which form the input of the forthcoming CCSD calculation.\newline
\indent $4)$ Next, the cluster amplitudes for the external part, 
i.e., $\hat T_1$ and $\hat T_2$, are calculated in the course of 
the DMRG-TCCSD scheme.\newline
\indent $5)$ As we discus in the next section, finding the 
optimal CAS, i.e., $k$-splitting, is a highly non-trivial problem, and at 
the present stage we can only present a solution that is considered as a 
heuristic approach in terms of rigorous mathematics. 
In practice, we repeat steps 2-4 for a large DMRG-truncation error as a 
function of $N/2<k<K$, thus we find local energy minima 
(see Fig.~\ref{fig:dmrg-k}) using a relatively 
cheap DMRG-TCCSD scheme. 
Around such a local minimum we perform more accurate DMRG-TCCSD calculations 
by lowering the DMRG-truncation error in order to refine the optimal $k$. 
We also monitor the maximum number of DMRG block states required to reach 
the {\it a priori} defined DMRG-error margin as a function of $k$. 
Since it can happen that several $k$ values lead to low error DMRG-TCCSD 
energies, while the computational effort increases significantly with 
increasing $k$ we select the optimal $k$ that leads to low DMRG-TCCSD 
energy but also minimizes the required DMRG block states. 
Using the optimal $k$ value we perform large-scale DMRG-TCCSD calculation 
using a relatively tight error bound for the DMRG-truncation error. 
\newline
\indent
We close this section with a brief summary of the numerically accessible error terms and relate them to equations presented in Sec.~IV.
Note that the error analysis in Sec.~IV is presented for a given $k$, thus here the $k$ dependence is also omitted.\newline
\indent
For a given $k$ split, the accuracy of $|\Psi^{{\rm CAS}}_{\rm DMRG}\rangle$ depends on the DMRG truncation error, $\delta\varepsilon_{\rm Tr}$.
As has been shown in Refs.~\cite{Legeza-1996,Legeza-2003a} the relative error, $\Delta E_{\rm rel} = (E^{{\rm CAS}}_{\rm DMRG(\delta\varepsilon_{\rm Tr})}-E^{{\rm CAS}}_{\rm FCI})/E^{{\rm CAS}}_{\rm FCI}$ is a liner function of $\delta\varepsilon_{\rm Tr}$ on a logarithmic scale. 
Therefore, extrapolation to the FCI limit can be carried out as a function of $\delta\varepsilon_{\rm Tr}$. 
In addition, the error term $\Delta\mathcal{E}_{\rm DMRG}(\delta\varepsilon_{\rm Tr})= E^{{\rm CAS}}_{\rm DMRG(\delta\varepsilon_{\rm Tr})}-E^{{\rm CAS}}_{\rm FCI}$ appearing in Eq.~\eqref{eq:WavefunctionError} can be controlled.\newline
\indent
Note that terms appearing in Eq.~\eqref{eq:SecondErrorTerm} and Eq.~\eqref{eq:WavefunctionError} include FCI solutions of the considered system. 
However, in special cases these can be well approximated as follows:
Considering a small enough system that is dynamically correlated, like the Nitrogen dimer near the equilibrium geometry with the here chosen basis set. 
The CI-coefficients are then extractable from the matrix product state representation of a wave function, e.g., $|\Psi_{\rm DMRG}^{\rm CAS}\rangle$ or $|\Psi_{\rm FCI}^{\rm CAS}\rangle$. 
Note that calculating all CI-coefficients scales exponentially with the size of the CAS.
However, since the system is dynamically correlated zeroth order, single and double excitation coefficients are sufficient.
Hence the error terms $|| \,|\Psi_{\rm DMRG}^{\rm CAS}\rangle-|\Psi_{\rm FCI}^{\rm CAS}\rangle||_{L^2}$ and $||(\hat S_{\rm FCI} - \hat S_{\rm DMRG(\delta\varepsilon_{\rm Tr})})|\Psi_{\rm HF}\rangle||$ in Eq.~\eqref{eq:SecondErrorTerm} and Eq.~\eqref{eq:WavefunctionError}, respectively, can be well approximated.
We remark that this exponential scaling with the CAS size also effects the computational costs of the CAS CI-triples, which are needed for an exact treatment of the TCCSD energy equation. 
However, investigations of the influence of the CAS CI-triples on the computed energies are left for future work.

\subsection*{B. Results and Discussion}
\label{sec:results}

In this section, we investigate the overall error dependence of DMRG-TCCSD as a function of $k$ and as a function of the DMRG-truncation error $\delta\varepsilon_{\rm Tr}$. 
For our numerical error study we perform steps 1-4 discussed in 
Sec.~V\,A for each $N/2<k<K$.
For each geometry $r=2.118\,a_0, 2.700\,a_0,$ and $3.600\,a_0$ we also carry out very high accuracy DMRG calculations on the full orbital space, i.e., by setting the truncation error to $\delta\varepsilon_{\rm Tr}=10^{-8}$ and $k=K$.
This data is used as a reference for the FCI solution.

%A CAS vektorok:
%/home/mmate/runs/DMRG-TCCSD/n2/4_redump/n2#rh2.118/CAS_orb.out
%CAS_orb = [ 6   7   8   9   5   4  17   3  14  16  15  10  13  12  11  21
%22  20  19  18  25  23  24  28  27  26   2   1 ];
%
%/home/mmate/runs/DMRG-TCCSD/n2/4_redump/n2#rh2.700/CAS_orb.out
%CAS_orb = [ 6   7   9   8   5  10   4   3  11  17  16  14  12  13  15  20
%19  22  21  18  25  24  23  27  26  28   2   1 ];
%
%/home/mmate/runs/DMRG-TCCSD/n2/4_redump/n2#rh3.600/CAS_orb.out
%CAS_orb = [ 7   6   8   9   5  10   4   3  13  11  12  14  16  15  17  19
%18  21  20  23  22  24  25  27  26  28   2   1 ];

\subsubsection*{1. An Entropy Study on the Full Orbital Space}
\label{sec:tcc-fullcas} 

\begin{figure}[htb]
\centerline{
\scalebox{0.4}{\includegraphics{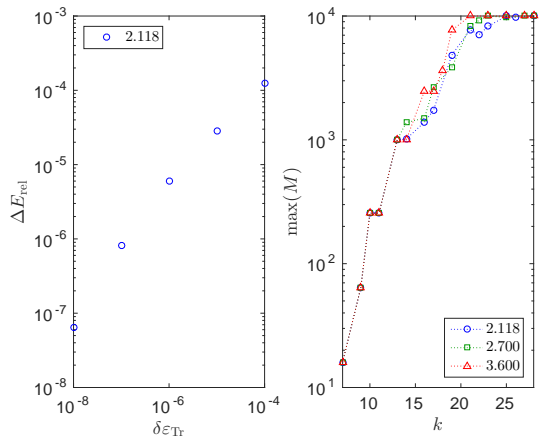}}
}
\caption{(a) Relative error of the ground-state energy as a function of the 
DMRG-truncation error on a logarithmic scale obtained for the full orbital 
space ($k=K$) with $r=2.118\,a_0$.
(b) Maximum number of block states as a function of $k$ for the {\it a priori} 
defined truncation error $\delta\varepsilon_{\rm Tr}=10^{-8}$ with 
$r=2.118\,a_0$ (blue), $2.700\,a_0$ (green), and $3.600\,a_0$ (red).
}
\label{fig:Erel-tre}
\end{figure}

We start our investigation by showing DMRG results for the full orbital space, i.e., the 
CAS is formed from $k=K=28$ orbitals, and for various fixed $M$ values and 
for $\delta\varepsilon_{\rm Tr}=10^{-8}$.  
In the latter case the maximum bond dimension was set to $M=10 000$.
%found to be around $M=10 000$.
%
In Fig.~\ref{fig:Erel-tre}~(a) we show the relative error of the ground-state 
energy as a function of the DMRG-truncation error on a logarithmic scale. 
For the FCI energy, $E_{\rm FCI}$, the CCSDTQPH reference energy is used given in Ref.~\cite{Chan-2004b}. 
It is visible that the relative error is a linear function of the truncation error on a logarithmic scale, thus extrapolation to the truncation free solution can be carried out according to Refs.~\cite{Legeza-1996,Legeza-2003a}.

\begin{figure}[htb]
\centerline{
\scalebox{0.4}{\includegraphics{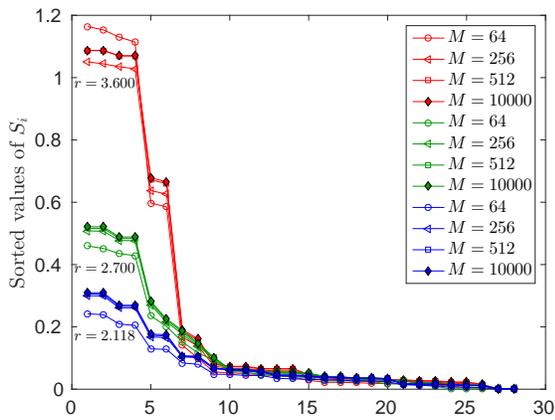}}
}
\caption{Single orbital entropy for $r=2.118\,a_0$ (blue), $2.700\,a_0$ (green), $3.600\,a_0$ (red) obtained for the full orbital space ($k=28$) with DMRG for fixed $M=64, 256, 512$ and for $\delta\varepsilon_{\rm Tr}=10^{-8}, M_{\rm max}=10 000$.
}
\label{fig:entropy-fullcas}
\end{figure}

\begin{figure}[htb]
\centerline{
\scalebox{0.4}{\includegraphics{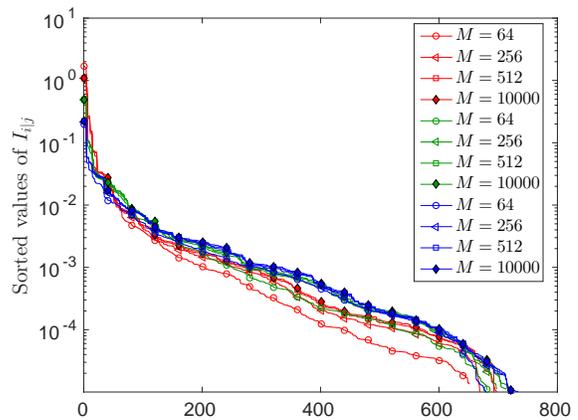}}
}
\caption{Mutual information 
for $r=2.118\,a_0$ (blue), $2.700\,a_0$ (green), $3.600\,a_0$ (red) obtained for the full orbital space ($k=28$) with DMRG for fixed $M=64, 256, 512$ and for $\delta\varepsilon_{\rm Tr}=10^{-8}, M_{\rm max} = 10 000$.
}
\label{fig:mutualinfo-fullcas}
\end{figure}

In Figs.~\ref{fig:entropy-fullcas} and \ref{fig:mutualinfo-fullcas} we present 
the sorted values of the single orbital entropy and of the mutual information 
obtained for fixed $M=64,256,512$ and with $\delta\varepsilon_{\rm Tr}=10^{-8}$ 
for the three geometries.
As can be seen in the figures the entropy profiles obtained with low-rank 
DMRG calculations already resemble the main characteristics of the exact 
profile ($M\simeq 10000$).
Therefore, orbitals with large single orbital entropies, also contributing 
to large matrix elements of $I_{i|j}$, can easily be identified from a 
low-rank computation.
The ordered orbital indices define the CAS-vector, and the CAS for the 
DMRG-TCCSD can be formed accordingly as discussed in Sec.~V\,A.

Taking a look at Fig.~\ref{fig:entropy-fullcas} it becomes apparent that 
$S_i$ shifts upwards for increasing $r$ indicating the higher 
contribution of static correlations for the stretched geometries. 
Similarly the first 50-100 matrix elements of $I_{i|j}$ also take larger 
values for larger $r$ while the exponential tail, corresponding to dynamic 
correlations, is less effected. 
The gap between large and small values of the orbital entropies gets larger 
and its position shifts rightward for larger $r$. Thus, for the stretched 
geometries more orbitals must be included in the CAS during the TCC scheme 
in order to determine the static correlations accurately.
We remark here, that the orbitals contributing to the high values 
of the single orbital entropy and mutual information matrix elements change 
for the different geometries according to chemical bond forming and breaking 
processes \cite{Boguslawski-2013}.

\subsubsection*{2. Numerical Investigation of the Error's $k$-dependence}
\label{sec:tcc-k} 

\begin{figure}[htb]
\centerline{
\scalebox{0.4}{\includegraphics{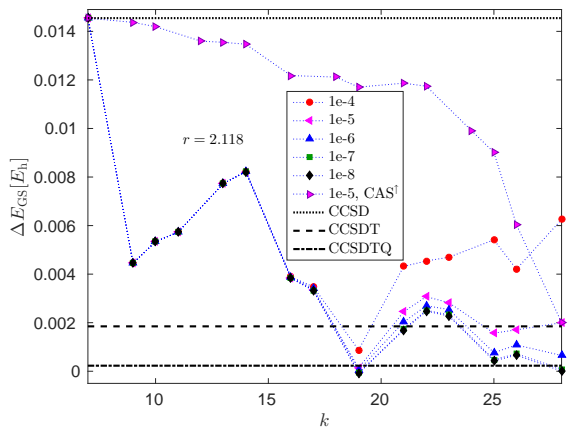}}
}
\caption{%(a)
Ground-state energy of the N$_2$ molecule near the equilibrium 
geometry, $r=2.118\,a_0$, obtained with DMRG-TCCSD for $7\le k\le 28$ 
and for various DMRG truncation errors $\delta\varepsilon_{\rm Tr}$.
The CCSD, CCSDT and CCSDTQ reference energies are shown by dotted, dashed and 
dashed-dotted lines, respectively. 
The DMRG energy with $\delta\varepsilon_{\rm Tr}=10^{-8}$ on the full space, i.e., $k = 28$, is taken as a reference for the FCI energy.  
For $\delta\varepsilon_{\rm Tr}=10^{-5}$ the CAS was additionally formed 
by taking $k$ orbitals according to increasing values of the single-orbital 
entropy values, i.e., inverse to the other CAS extensions.
This is labeled by CAS$^\uparrow$. 
}
\label{fig:dmrg-k}
\end{figure}

In order to obtain $|\Psi^*\rangle$ in the FCI limit, we perform
high-accuracy DMRG calculations with $\delta\varepsilon_{\rm Tr}=10^{-8}$.
The CAS was formed by including all Hartee--Fock orbitals and its size was 
increased systematically by including orbitals with the largest entropies 
according to the CAS vector.
Orbitals with degenerate single orbital entropies, due to symmetry 
considerations, are added to the CAS at the same time. 
Thus there are some missing $k$ points in the following figures.
For each restricted CAS we carry our the usual optimization steps of a 
DMRG scheme as 
discussed in Sec.~V\,A, with low bond dimension followed by a 
high-accuracy calculation with $\delta\varepsilon_{\rm Tr}=10^{-8}$ using 
eight sweeps \cite{Szalay-2015}. 
Our DMRG ground-state energies for $7<k<28$ together with the CCSD 
(corresponding to a DMRG-TCCSD calculation where $k=N/2=7$), 
and CCSDTQ reference energies, are shown in Fig.~\ref{fig:dmrg-k} 
near the equilibrium bond length, $r=2.118\,a_0$.  
The single-reference coupled cluster calculations were performed in NWChem \cite{VALIEV20101477}, we employed the cc-pVDZ basis set in the spherical representation.
For $k=K=28$ the CCSDTQPH energy was taken as a reference for the FCI 
energy \cite{Chan-2004b}.

\begin{figure}[htb]
\centerline{
\scalebox{0.4}{\includegraphics{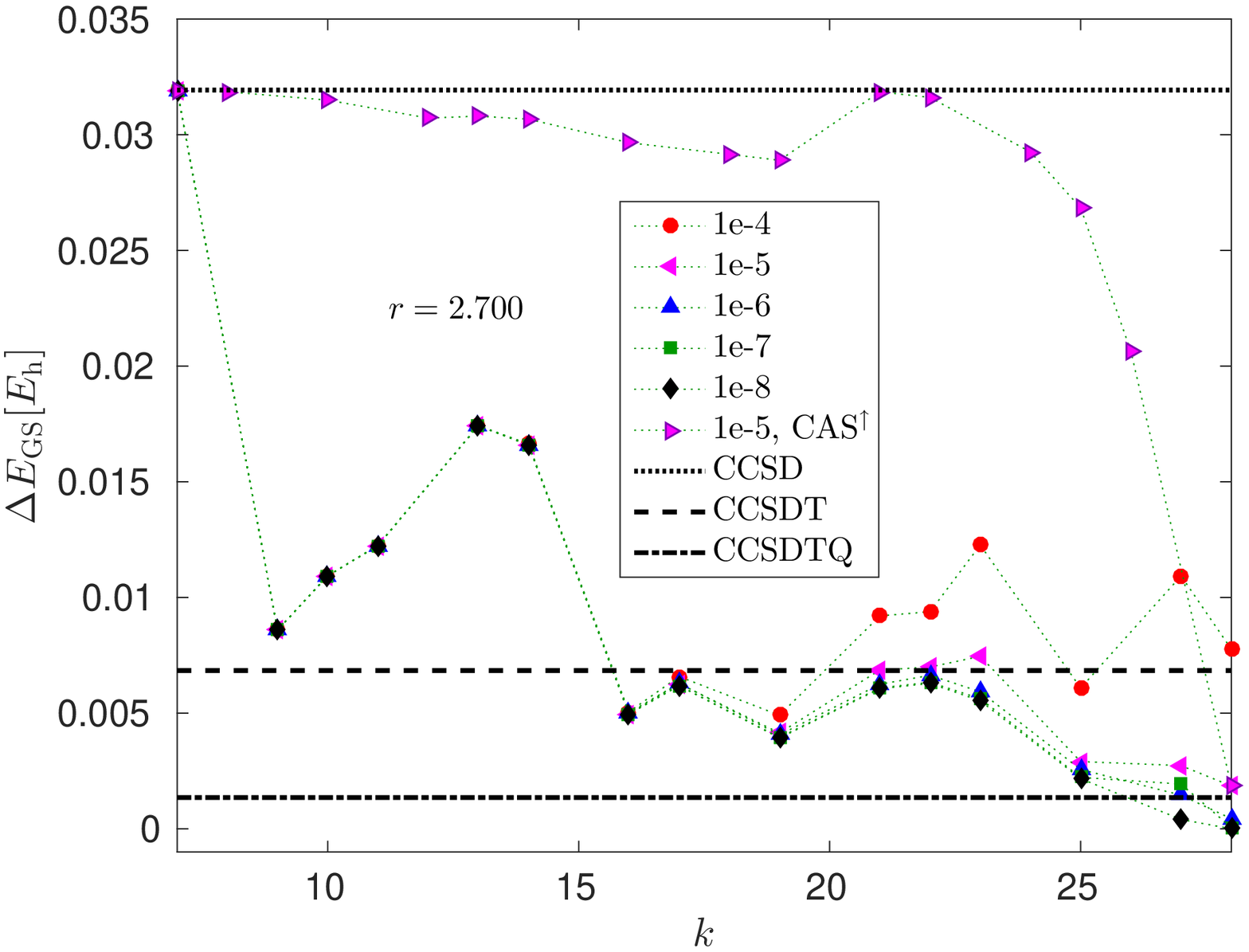}}
}
\caption{
Ground-state energy of the N$_2$ molecule with bond length $r=2.7\,a_0$, obtained with DMRG-TCCSD for $7\le k\le 28$ 
and for various DMRG truncation errors $\delta\varepsilon_{\rm Tr}$.
The CCSD, CCSDT and CCSDTQ reference energies are shown by dotted, dashed and 
dashed-dotted lines, respectively. 
The DMRG energy with $\delta\varepsilon_{\rm Tr}=10^{-8}$ on the full space, i.e., $k = 28$, is taken as a reference for the FCI energy.  
For $\delta\varepsilon_{\mathrm{Tr}} = 10^{-5}$ the CAS was additionally formed by taking $k$ orbitals according to increasing values of the single-orbital entropy, i.e., inverse to the other CAS extensions. 
This is labeled by CAS$^{\uparrow}$.
}
\label{fig:dmrg-k-2.7}
\end{figure}

\begin{figure}[htb]
\centerline{
\scalebox{0.4}{\includegraphics{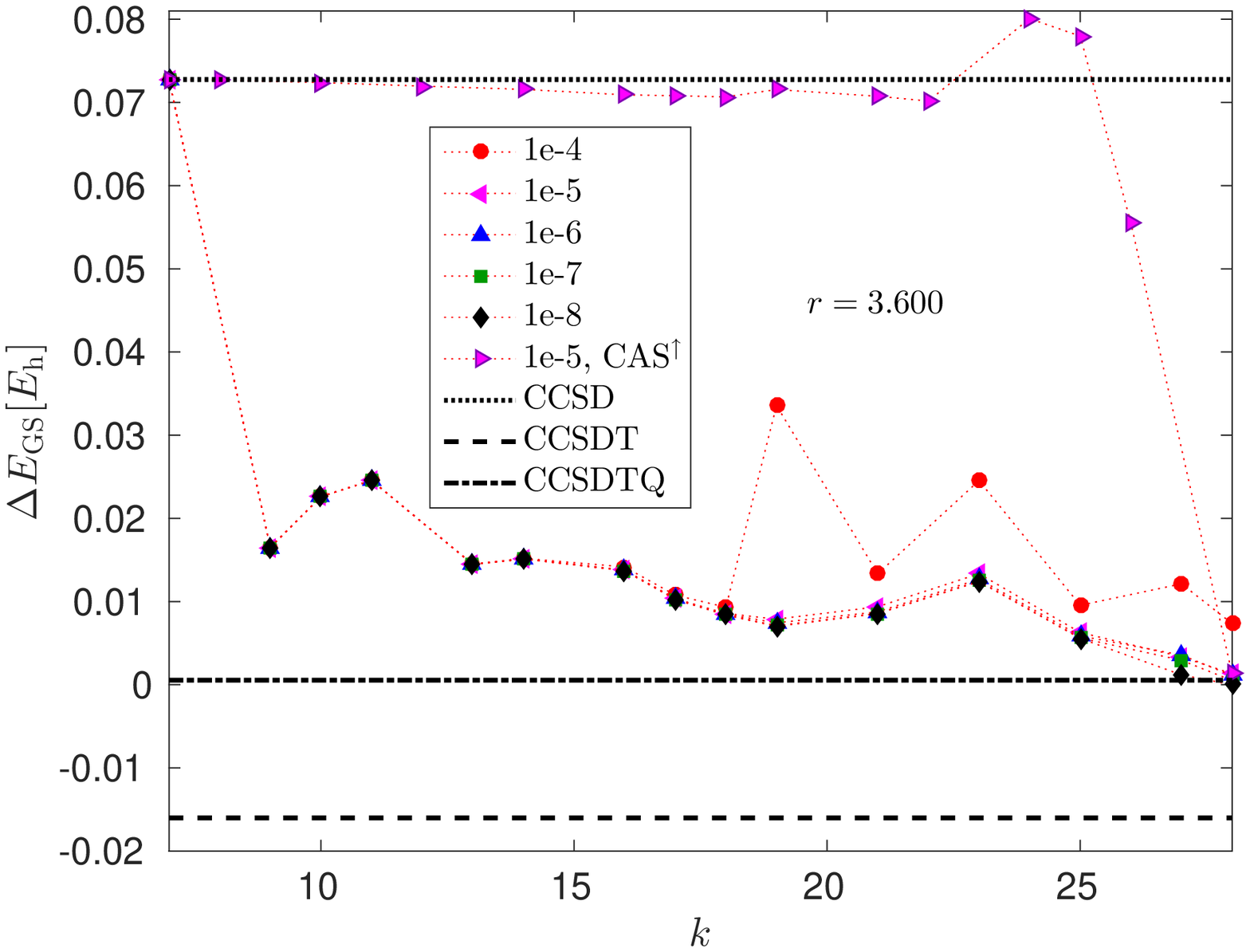}}
}
\caption{
	Ground-state energy of the N$_2$ molecule with bond length $r=3.6\,a_0$, obtained with DMRG-TCCSD for $7\le k\le 28$ 
	and for various DMRG truncation errors $\delta\varepsilon_{\rm Tr}$.
	The CCSD, CCSDT and CCSDTQ reference energies are shown by dotted, dashed and 
	dashed-dotted lines, respectively. 
	The DMRG energy with $\delta\varepsilon_{\rm Tr}=10^{-8}$ on the full space, i.e., $k = 28$, is taken as a reference for the FCI energy.  
	For $\delta\varepsilon_{\mathrm{Tr}} = 10^{-5}$ the CAS was additionally formed by taking $k$ orbitals according to increasing values of the single-orbital entropy, i.e., inverse to the other CAS extensions. 
	This is labeled by CAS$^{\uparrow}$.
}
\label{fig:dmrg-k-3.6}
\end{figure}

The DMRG energy starts from the Hartree--Fock energy for $k=7$ and decreases monotonically with increasing $k$ until the full orbital solution with $k=28$  is reached.
It is remarkable, however, that the DMRG-TCCSD energy is significantly below the CCSD energy for all CAS choices, even for a very small $k=9$.
The error, however, shows an irregular behavior taking small values for several different $k$-s.
This is due to the fact that the DMRG-TCCSD approach suffers from a 
methodological 
error, i.e., certain fraction of the correlations are lost, since the CAS is 
frozen in the CCSD correction.
This supports the hypothesis of a $k$-dependent constant as discussed in 
Sec.~IV\,D. 
Therefore, whether orbital $k$ is part of the CAS or external part provides 
a different methodological error.
This is clearly seen as the error increases between $k=10$ and $k=15$ 
although the CAS covers more of the system's static correlation with 
increasing $k$.
This is investigated in more detail in Sec.~V\,B\,4. \newline
\indent Since several $k$-splits lead to small DMRG-TCCSD errors, the optimal 
$k$ value from the computational point of view, is determined not only by 
the error minimum but also by the minimal computational time, i.e., 
we need to take the computational requirements of the DMRG into account.
Note that the size of the DMRG block states contributes significantly to 
the computational cost of the DMRG calculation.
The connection of the block size to the CAS choice is shown in 
Fig.~\ref{fig:Erel-tre}~(b), where the maximal number of DMRG block 
states is depicted as a function of $k$ for 
the {\it a priori} defined truncation error margin $\delta\varepsilon_{\rm Tr}=10^{-8}$.
Note that $\max(M)$ increases rapidly for $10<k<20$. 
The optimal CAS is therefore chosen such that the DMRG block states are not 
too large and the DMRG-TCCSD provides a low error, i.e., is a local minimum 
in the residual with respect to $k$.\newline
\indent It is important to note that based on Fig.~\ref{fig:dmrg-k} the DMRG-TCCSD 
energy got very close to, or even dropped below, the CCSDT energy for 
several $k$ values. 
Since close to the equilibrium geometry the wave function is 
dominated by a single reference character, it is expected that DMRG-TCCSD 
leads to even more robust improvements for the stretched geometries, i.e., 
when the multi-reference character of the wave function is more pronounced.
Our results for the stretched geometries, $r=2.700\,a_0$ and $3.600\,a_0$, 
are shown in Figs.~\ref{fig:entropy-fullcas},~\ref{fig:mutualinfo-fullcas}~\ref{fig:dmrg-k-2.7} 
and~\ref{fig:dmrg-k-3.6}.
As mentioned in Sec.~V\,B\,1, for larger $r$ values static 
correlations gain importance signaled by the increase in the single orbital 
entropy in Fig.~\ref{fig:entropy-fullcas}.
Thus the multi-reference character
of the wave function becomes apparent through the entropy profiles.
According to Fig.~\ref{fig:dmrg-k-2.7} the DMRG-TCCSD energy for all $k>7$ 
values is again below the CCSD computation and for $k>15$ it is even below 
the CCSDT reference energy.
For $r=3.600\,a_0$ the CC computation fluctuates with increasing excitation 
ranks and CCSDT is even far below the FCI reference energy, revealing 
the variational breakdown of the single-reference CC method for multi-reference problems. 
In contrast to this, the DMRG-TCCSD energy is again below the CCSD energy for 
all $k>7$, but above the the CCSDT energy.
The error furthermore shows a local minimum around $k=19$.
For the stretched geometries static correlations are more pronounced, there are more orbitals with large entropies, thus the maximum number of DMRG block states increases more rapidly with $k$ compared to the situation near the equilibrium geometry (see Fig.~\ref{fig:Erel-tre}~(b)). 
Thus obtaining an error margin within one milli-Hartree for $k=19\ll28$ leads to a significant save in computational time and resources.

\subsubsection*{3. Effect of $\delta\varepsilon_{\rm Tr}$ on the DMRG-TCCSD}
\label{sec:dmrg-tre} 

In practice, we do not intend to carry out DMRG calculations in the FCI 
limit, thus usually a larger truncation error is used. 
Therefore, we have repeated our calculations for larger truncation errors 
in the range of $10^{-4}$ and $10^{-7}$.
Our results are shown in Figs.~\ref{fig:dmrg-k}, \ref{fig:dmrg-k-2.7}, 
and \ref{fig:dmrg-k-3.6}.
For small $k$ the DMRG solution basically provides the Full-CI limit since the a priori set minimum number of block states $M_{\rm min}\simeq 64$ already leads to a very low truncation error.
Therefore, the error of the DMRG-TCCSD is dominated by the methodological error.
For $k>15$ the effect of the DMRG truncation error becomes visible and for large $k$ the overall error is basically determined by the DMRG solution.
For larger $\delta\varepsilon_{\rm Tr}$ between $10^{-4}$ and $10^{-5}$ the DMRG-TCCSD error 
shows a minimum with respect to $k$. 
This is exactly the expected trend, since the CCSD method fails to capture 
static correlation while DMRG requires large bond dimension to recover 
dynamic correlations, i.e., a low truncation error threshold. 
In addition, the error minima for different truncation error thresholds 
$\delta\varepsilon_{\rm Tr}$ happen to be around the same $k$ values. 
This has an important practical consequence: the optimal $k$-split can be 
determined by performing cheap DMRG-TCCSD calculations using large DMRG 
truncation error threshold as a function of $k$. \newline
\indent  
The figures furthermore indicate that $\Delta E_{\rm GS}$ has a high peak 
for $9<k<16$.
This can be explained by the splitting of the FCI space since this yields 
that the correlation from external orbitals with CAS orbitals is ignored. 
Thus we also performed calculations for $\delta\varepsilon_{\rm Tr}=10^{-5}$ using a CAS formed by taking $k$ orbitals according to increasing values of the single orbital entropy values in order to demonstrate the importance of the CAS extension. 
The corresponding error profile as a function of $k$ near the equilibrium 
geometry is shown in Fig.~\ref{fig:dmrg-k} labeled by CAS$^\uparrow$.
As expected, the improvement of DMRG-TCCSD is marginal compared to CCSD up to a very large $k\simeq 23$ split since $\psi^{\rm CAS}_{\rm DMRG}$ differs only marginally from $\psi_{\rm HF}$.

\subsubsection*{4. Numerical Investigation on CAS-ext correlations}
\label{sec:cas-ext} 

Taking another look at Fig.~\ref{fig:entropy-fullcas}, we can confirm that 
already for small $k$ values the most important orbitals, i.e., those with 
the largest entropies, are included in the CAS.
\begin{figure}[htb]
\centerline{
\scalebox{0.45}{\includegraphics{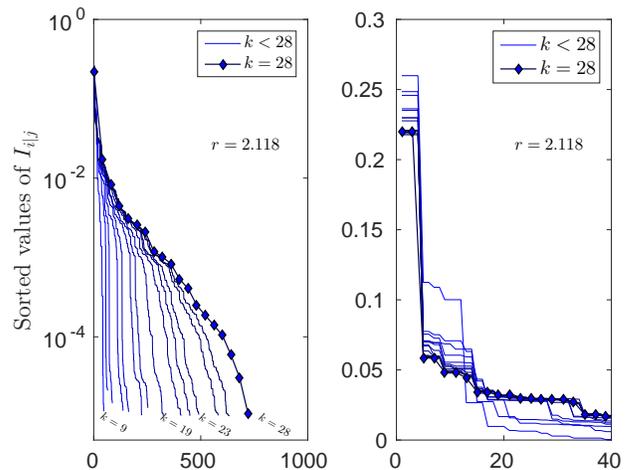}}
}
\caption{(a) Sorted values of the mutual information obtained 
by DMRG($k$) for $9\leq k\leq 28$ on 
a semi-logarithmic scale for N$_2$ at $r=2.118\,a_0$. 
(b) 
Sorted 40 largest matrix elements of the mutual information obtained 
by DMRG($k$) for $9\leq k\leq 28$ on 
a lin-lin scale for N$_2$ at $r=2.118\,a_0$
}
\label{fig:Idecay-k}
\end{figure}
In Fig.~\ref{fig:Idecay-k} the sorted values of the mutual information obtained 
by DMRG($k$) for $9\leq k\leq 28$ is shown on a semi-logarithmic scale. 
It is apparent from the figure that the largest values of $I_{i|j}$ change only slightly with increasing $k$, thus static correlations are basically included for all restricted CAS. %
The exponential tail of $I_{i|j}$ corresponding to dynamic correlations, however, becomes more visible only for larger $k$ values.
We conclude, for a given $k$ split the DMRG method computes the static correlations efficiently and the missing tail of the mutual information with respect to the full orbital space ($k=28$) calculation is captured by the TCC scheme.\newline  
\indent Correlations between the CAS and external parts can also be simulated 
by a DMRG calculation on the full orbital space using an orbital ordering 
according to the CAS-vector. 
In this case, the DMRG left block can be considered as the CAS and the right 
block as the external part.
For a pure target state, for example, the ground state, the correlations between the CAS and external part is measured by the block entropy, 
$S(\rho_{{\rm CAS}(k)})$ as a function of $k$. 
Here $\rho_{{\rm CAS}(k)}$ is formed by a partial trace on the external part of 
$|\Psi^{\rm FCI}_{\rm DMRG}\rangle$. 
The block entropy is shown in Fig.~\ref{fig:error2.118}~(a).
%
%\begin{figure}[htb]
%\centerline{
%\scalebox{0.45}{\includegraphics{{figs/n2_sblock-k_2.118}.eps}}
%}
%\caption{Block entropy, $S(\rho_{{\rm CAS}(k)})$ as a function of $k$ for $r=2.118$
%ordering orbitals along the DMRG chain according to 
%the same CAS and CAS$^\uparrow$ vectors as used in Fig.~\ref{fig:dmrg-k}.}
%\label{fig:s-k-2.118}
%\end{figure}
%
The block entropy decays monotonically for $k>7$, i.e, the correlations between the CAS and the external part vanish with increasing $k$. 
In contrast to this, when an ordering according to CAS$^\uparrow$ is used the 
correlation between CAS and external part remains always strong, i.e., some of 
the highly correlated orbitals are distributed among the CAS and the external 
part. 
Nevertheless, both curves are smooth and they cannot explain the error 
profile shown in Fig.~\ref{fig:dmrg-k}.

\subsubsection*{5. Numerical Values for the Amplitude Error Analysis}
\label{sec:error-k-amplitudes}

\begin{figure}[htb]
	\centerline{
		\scalebox{0.48}{\includegraphics{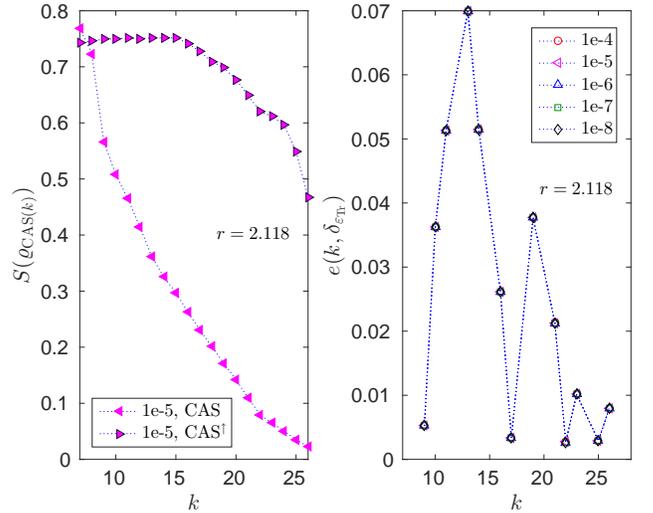}}
	}
	\caption{(a) Block entropy, $S(\rho_{{\rm CAS}(k)})$ as a function of $k$ for $r=2.118$
		ordering orbitals along the DMRG chain according to 
		the same CAS and CAS$^\uparrow$ vectors as used in Fig.~\ref{fig:dmrg-k}. (b) $e(k,\delta_{\varepsilon_{\rm Tr}})$ as a function of $k$ of the Nitrogen dimer near the equilibrium bond length for DMRG truncation error thresholds $\delta_{\varepsilon_\mathrm{Tr}}$ between $10^{-4}$ and $10^{-8}$.
	}
	\label{fig:error2.118}
\end{figure}

Since correlation analysis based on the entropy functions cannot
reveal the error profile shown in Fig.~\ref{fig:dmrg-k}, 
here we reinvestagte the error behavior as a function of 
$N/2 \leq k \leq K$ but in terms of the CC amlitudes.
Therefore, we also present a more detailed description of 
Eq.~\eqref{eq:error-k} in 
Sec.~IV which includes the following
terms:
\begin{equation}
\label{eq:error-k-long}
\begin{aligned}
e(k,\delta_{\varepsilon_\mathrm{Tr}}) &=  
\sum_{\substack{\mu:\\ \abs{\mu}=1}} \big( t_{\mathrm{CCSD}}(k,\delta_{\varepsilon_\mathrm{Tr}}) \big)^2_{\mu}\\
&\quad+
\sum_{\substack{\mu:\\ \abs{\mu}=1,2}} \Big[ 
\big( t_k^* - t_{\mathrm{CCSD}}(k,\delta_{\varepsilon_\mathrm{Tr}}) \big)^2_{\mu}\\
&\quad+
\big( s_k^* - s_{\mathrm{DMRG}}(k,\delta_{\varepsilon_\mathrm{Tr}}) \big)^2_{\mu} 
\Big].
\end{aligned}
\end{equation}
%\begin{eqnarray}
% e(k,\delta_{\varepsilon_\mathrm{Tr}}) &=&  
% \sum_{\substack{\mu:\\ \abs{\mu}=1}} \big( t_{\mathrm{CCSD}}(k,\delta_{\varepsilon_\mathrm{Tr}}) \big)^2_{\mu} \nonumber\\
%&+&
% \sum_{\substack{\mu:\\ \abs{\mu}=1,2}} \Big[ 
%  \big( t_k^* - t_{\mathrm{CCSD}}(k,\delta_{\varepsilon_\mathrm{Tr}}) \big)^2_{\mu} \nonumber\\
%&+& 
%  \big( s_k^* - s_{\mathrm{DMRG}}(k,\delta_{\varepsilon_\mathrm{Tr}}) \big)^2_{\mu} 
% \Big].
%\label{eq:error-k-long}
%\end{eqnarray}
Here the \emph{valid index-pairs} are $\mu = (\bm{i},\bm{a})$, with 
$\bm{i}=(i_1,\ldots,i_n) \in \set{1,\ldots,N/2}^n $,  and 
$\bm{a}=(a_1,\ldots,a_n) \in \set{N/2+1,\ldots,K}^n$. 
The excitation rank is given by $\abs{\mu} = n$ where $n=1$ stands for
 singles, $n=2$ for doubles, and so on.
The $\mu$-s are the labels of excitation operators
$\hat{\tau}_{i}^{a} := \hat{a}_{a}^{\dagger} \hat{a}_{i}$, 
and $\hat{\tau}_{i_1,\ldots,i_n}^{a_1,\ldots,a_n} := \hat{\tau}_{i_n}^{a_n} \ldots \hat{\tau}_{i_1}^{a_1} $. 
The corresponding amplitudes are given as $t_{i_1,\ldots,i_n}^{a_1,\ldots,a_n}$.
For invalid index-pairs, i.e., index-pairs that are out of range, 
the amplitudes are always zero.
The various amplitudes appering in Eq.~\eqref{eq:error-k-long} are calculated
according to the following rules:\newline
\indent 1) The $s_k^*$: amplitudes in the CAS($k$) obtained by DMRG($\delta^{\mathrm{*}}_{\varepsilon_\mathrm{Tr}}=10^{-8}$) solution (represented by CI coefficients $c^*$) for CAS($K$), 
\begin{equation}
\label{eq:ci2cc}
\begin{aligned}
(s_k^*)_i^a &= \frac{c^{*a}_i}{c^*_0}, \\%\qquad
(s_k^*)_{i_1,i_2}^{a_1,a_2} &= \frac{c^{*a_1,a_2}_{i_1,i_2}}{c^*_0} - \frac{c^{*a_1}_{i_1} c^{*a_2}_{i_2} - c^{*a_2}_{i_1} c^{*a_1}_{i_2} }{c^{*2}_0} 
\end{aligned}
\end{equation}
where $i,i_1,i_2 \in \set{1,\ldots,N/2}$ and $a,a_1,a_2 \in \set{N/2+1,\ldots,k}$.\newline
\indent 2) The $t_k^*$: amplitudes not in the CAS($k$) obtained from the 
DMRG($\delta^{\mathrm{*}}_{\varepsilon_\mathrm{Tr}}=10^{-8}$) solution 
(represented by CI coefficients $c^*$) for CAS($K$) projected onto CAS($k$), i.e., the complement (with respect to valid index-pairs) of $s_k^*$.\newline
\indent 3) The $s_{\mathrm{DMRG}}(k,\delta_{\varepsilon_\mathrm{Tr}})$ amplitudes 
in the CAS($k$) are obtained by the DMRG($\delta_{\varepsilon_\mathrm{Tr}}$) 
solution (represented by CI coefficients $c$) for CAS($k$).
The amplitudes $s_{\mathrm{DMRG}}(k,\delta_{\varepsilon_\mathrm{Tr}})_{i}^{a}, s_{\mathrm{DMRG}}(k,\delta_{\varepsilon_\mathrm{Tr}})_{i_1,i_2}^{a_1,a_2}$ are the same as Eq.~\eqref{eq:ci2cc}, but with $c^* \to c$, where $i,i_1,i_2 \in \set{1,\ldots,N/2}$ and $a,a_1,a_2 \in \set{N/2+1,\ldots,k}$.\newline
\indent 4) The $t_{\mathrm{CCSD}}(k,\delta_{\varepsilon_\mathrm{Tr}})$: amplitudes not in the CAS($k$) obtained by TCCSD, i.e., the complement (with respect to valid index-pairs) of $s_{\mathrm{DMRG}}(k,\delta_{\varepsilon_\mathrm{Tr}})$. \newline
\indent 
In Fig.~\ref{fig:error2.118}~(b) we show the error $e(k,\delta_{\varepsilon_{\rm Tr}})$ as a function of $k$ of the Nitrogen dimer near the equilibrium bond length.
Note that the quantitative behavior is quite robust with respect to the bond dimension since the values only differ marginally.
We emphasize that the error contribution in Fig.~\ref{fig:error2.118} is dominated by second term in Eq.~\eqref{eq:error-k-long} since this is an order of magnitude larger than the contribution from the first and third terms in Eq.~\eqref{eq:error-k-long}, respectively. 
The first term in Eq.~\eqref{eq:error-k-long} is furthermore related to the usual T1 diagnostic in CC \cite{Lee1989},
so it is not a surprise that a small value, $\sim10^{-3}$, was found.
Comparing this error profile to the one shown in Fig.~\ref{fig:dmrg-k} we can understand the irregular behavior and the peak in the error in $\Delta E_{\rm GS}$ between $k=9$ and $17$, and the other peaks for $k>17$  but the error minimum found for $k=19$ remains unexplained.
Furthermore, we can conclude from Fig.~\ref{fig:error2.118}~(b) that the quotient $\Delta E_{\rm GS}(k)/e(k,\delta_{\varepsilon_{\rm Tr}})$ is not constant. 
This indicates that the constants involved in Sec.~IV in particular the constant in Eq.~\eqref{eq:error-k} in Sec.~IV\,D is indeed $k$ dependent.

\section*{VI. Conclusion}
\label{Sec:Conclusion}

In this article we presented a fundamental study of the DMRG-TCCSD method.
We showed that, unlike the traditional single-reference CC method, the linked and unlinked DMRG-TCC equations are in general not equivalent. 
Furthermore, we showed energy size extensivity of the TCC, DMRG-TCC and DMRG-TCCSD method and gave a proof that CAS excitations higher than order three do not enter the TCC energy expression.\newline
\indent
In addition to these computational properties of the DMRG-TCCSD method we presented the mathematical error analysis performed in Ref.~\cite{faulstich2018analysis} from a quantum chemistry perspective.
We showed local uniqueness and quasi optimality of DMRG-TCC solutions, and highlighted the importance of the CAS-ext gap -- a spectral gap assumption allowing to perform the analysis presented here. 
Furthermore, we presented a quadratic {\it a priori} error estimate for the DMRG-TCC method, which aligns the error behavior of the DMRG-TCC method with variational methods except for the upper bound condition.
We emphasize that the DMRG-TCC solution depends strongly on the CAS choice. 
Throughout the analysis we neglected this dependence as we assumed an optimal CAS choice as indicated in Sec.~IV\,A.
The explicit consideration of this dependence in the performed error analysis carries many mathematical challenges, which are part of our current research. 
Therefore, we extended this work with a numerical study of the k-dependence of the DMRG-TCCSD error.\newline
\indent
We presented computational data of the single-site entropy and the mutual information that are used to choose the CAS. 
Our computations showed that these properties are qualitatively very robust, i.e., their qualitative behavior is well represent by means of a low-rank approximation, which is a computational benefit. 
The numerical investigation of the $k$-dependence of the DMRG-TCCSD error revealed that the predicted trend in Sec.~IV\,A is correct. 
We can clearly see that the error indeed first decays (for $7\leq k\leq 9$) and then increases again (for $25\leq k\leq 28$) for low-rank approximations, i.e., 1e-4 respectively 1e-5.
This aligns with the theoretical prediction based on the properties of the DMRG and single reference CC method.
Additional to this general trend, the error shows oscillations. 
A first hypothesis is that this behavior is related to the ignored correlations in the transition $k\to k+1$. 
However, this was not able to be proven so far using entropy based measures but a similar irregular behavior can be detected by a cluster amplitude error analysis.
The error minimum found for the DMRG-TCCSD method, however, was not able to be proven within this article and is left for future work.
%
%\comment{I disagree, for M~10 000 this cannot be. delete: A second hypothesis is that the DMRG did not converge to the global minimum for certain $k$-values. 
%%
%This is possible since the increasing basis set changes the tensor manifold which can lead to a loss of monotonicity. 
%%
%However, also this hypothesis was not able to be proven within this article and remains for future works.} 
%
An important feature that we would like to highlight here is that a small CAS ($k=9$) yields a significant improvement of the energy, and that the energies for all three geometries and all CAS choices outrun the single-reference CC method.
Although the computational costs of the DMRG-TCCSD method exceed
the costs of the CCSD method, this leads to a computational drawback of the
method only if the treatment of large CAS becomes necessary.\newline
\indent In addition, the DMRG-TCCSD method avoids the numerical breakdown of the CC approach even for multi-reference (strongly correlated) systems and, using concepts of quantum information theory, allows an efficient black-box implementation.
The numerical investigations showed also that the constants involved in the error estimation are most likely $k$ dependent. 
This stresses the importance of further mathematical work to include this dependence explicitly in the analysis.\newline
\indent
Since the numerical error study showed a significant improvement for small CAS, we suspect the DMRG-TCCSD method to be of great use for larger systems with many strongly correlated orbitals as well as a many dynamically correlated orbitals \cite{veis2016coupled,veis2018intricate}. 
The oscillatory behavior of the error, however, remains unexplained at this point. 
Despite the unknown reason of this behavior, we note that the error minima are fairly robust with respect to the bond dimension. 
Hence, the DMRG-TCCSD method can be extended with a screening process using low bond-dimension approximations to detect possible error minima. 
In addition, our analysis is basis dependent.
Thus there is a need for further investigations based on, for example, fermionic mode transformation \cite{krumnow2016fermionic}, and investigations of the influence of the CAS CI-triples on the computed energies.
All these tasks, however, remain for future work.

\section*{Acknowledgements}
This work has received funding from the {\it Research Council of Norway} (RCN) under
CoE Grant No. 262695 (Hylleraas Centre for Quantum Molecular Sciences), from ERC-STG-2014
under grant No. 639508, from the {\it Hungarian National Research, Development and Innovation Office}
(NKFIH) through Grant No. K120569, from the {\it Hungarian Quantum Technology
National Excellence Program} (Project No. 2017-1.2.1-NKP-2017-00001), from the {\it Czech
Science Foundation} (grants no. 16-12052S, 18-18940Y, and 18-24563S), and the {\it Czech Ministry of Education, Youth and Sports} (project no. LTAUSA17033).
FMF, AL, \"OL, MAC  are grateful for
the mutual hospitality received during their visits at the Wigner Research Center for Physics in Budapest and the Hylleraas Centre for Quantum Molecular Sciences.
\"OL and JP acknowledge useful discussions with Marcel Nooijen.

\bibliographystyle{unsrt}
\bibliography{lib}
\end{document}